\documentclass[preprint,12pt,nofootinbib]{revtex4-1}

\usepackage{graphicx}\graphicspath{%
{graph/}}%
\usepackage{amssymb,amsmath,amsfonts}
\usepackage{time}
\usepackage{epstopdf}
\usepackage{hyperref}
\hypersetup{
  colorlinks,
  citecolor=magenta,
  linkcolor=blue}
\usepackage[T2A,T1]{fontenc} 
\usepackage[utf8]{inputenc} 
\usepackage{txfonts} 

\newcommand{\p}{{\partial}}

\renewcommand{\Re}{{\rm Re\,}}
\renewcommand{\Im}{{\rm Im\,}}
\renewcommand{\vec}[1]{\textnormal{\boldmath$#1$}}
\begin{document}

\title{Analytical theory of coherent synchrotron radiation wakefield of short bunches shielded by conducting parallel plates \footnote[1]{Work supported by the Department of Energy,
contract DE-AC03-76SF00515} }

\author{Gennady Stupakov}
\affiliation{SLAC National Accelerator Laboratory,
Menlo Park, CA 94025}
\author{Demin Zhou}
\affiliation{KEK, 1-1 Oho, Tsukuba, Ibaraki 305-0801, Japan}

\begin{abstract}
We develop a general model of coherent synchrotron radiation (CSR) impedance with shielding provided by two parallel conducting plates. This model allows us to easily reproduce all previously known analytical CSR wakes and to expand the analysis to the situations not explored before. It reduces calculations of the impedance to taking integrals along the trajectory of the beam. New analytical results are derived for  the radiation impedance with shielding for the following orbits: a kink, a bending magnet, a wiggler of finite length, and an infinitely long wiggler. All our formulas are benchmarked agains numerical simulations with the CSRZ computer code.
\end{abstract}

\maketitle

%
\section{Introduction}\label{sec:I}
%

Coherent synchrotron radiation (CSR) of short relativistic beams and its effect on beam dynamics in modern accelerators has been an area of active research for more than two decades. Various methods of calculation of the CSR wakefield were proposed in the literature. One of the first, and the simplest, approaches~\cite{derbenev95rss} treats the beam as having negligible transverse dimensions (a \emph{line charge} model) and neglects the effect of the walls of the vacuum chamber (the \emph{free-space} CSR wakefield). While the results of this model are applicable for relatively long magnets, the model is extremely useful for crude and quick estimates of the CSR effects in the system. A more complicated model~\cite{murphy97kg} takes into account the shielding effect of the vacuum chamber by approximating the metal walls by two parallel conducting plates located on the opposite sides of the beam circular orbit. Even more sophisticated approaches of Refs.~\cite{warnock90m,ng90} solve the synchrotron radiation and find the beam impedance in a toroidal vacuum chamber of rectangular cross section. 

Analyses of Refs.~\cite{derbenev95rss,murphy97kg,warnock90m,ng90} are limited to a circular trajectory of the beam. An important next step has been made in Ref.~\cite{saldin97sy}, where the authors considered a bending magnet of finite length and calculated the CSR wakefield for a trajectory consisting of an arc of a circle with incoming and exiting straight lines. This model made it possible to study effects of CSR radiation in bunch compressors of modern x-ray free electron lasers, where short bending magnets are separated by long drift sections. A simplified version~\cite{stupakov02e} of the CSR wake~\cite{saldin97sy} valid in the limit $v=c$ ($v$ is the particle velocity and $c$ is the speed of light) is implemented in the computer code elegant~\cite{borland00}. In a subsequent paper~\cite{saldin98sy} the authors of~\cite{saldin97sy} applied the same method to the calculation of the CSR wake in an infinitely long undulator in free space. Modification of the CSR wakefield derived in~\cite{saldin98sy} for the limit $v= c$ was carried out in~\cite{wu03rs}. 

In addition to various analytical approaches to the problem of CSR wakefield mentioned above, there have been a consistent effort to develop numerical algorithms for computer codes that calculate the wake in practically realistic situations. A good review of such codes can be found in Ref.~\cite{bassi06etal} with some latest additions to the list in Refs.~\cite{li08,mayes09,PhysRevSTAB.12.040703}. While these codes are indispensable in the design of accelerators, it is our opinion, that they do not eliminate the need for further development of new analytical tools that allow for a quick evaluation of the CSR effects in various conditions. In addition, the analytical approach usually provide the scalings of the strength of the effect and a better understanding of the mechanisms that cause the wakes. This, in turn, often allows find a solution that mitigates the adverse effect of the CSR wakefields.

In this paper we develop a general model of CSR impedance with shielding provided by two parallel conducting plates. This model reproduces all previous examples known from the literature and expands the analysis to the situations not explored before. It reduces  calculations of the impedance to taking integrals along the trajectory of the beam. These integrals can often be easily computed numerically with the help of Matlab or Mathematica.

The paper is organized as follows. In Section~\ref{sec:II}, starting from the retarded potentials of a relativistic beam in free space, we derive an expression for the radiation impedance in terms of integrals taken along the beam orbit. In Section~\ref{sec:III}, this expression is generalized to the case of shielding with parallel conducting plates. In Section~\ref{sec:IV} we give a brief description of the computer code CSRZ that we use for benchmarking our analytical results. In Section~\ref{sec:V} we reproduce some known results: the CSR impedance of a circular orbit in free space and with shielding, and the impedance of infinitely long wiggler in free space. In Section~\ref{sec:VI} we derive the impedance of a kink, that is an orbit consisting of two straight lines at a small angle. In Sections~\ref{sec:VII}, \ref{sec:VIII} and~\ref{sec:IX}   we derive the impedance of a bending magnet of finite length, a finite length wiggler and an infinite wiggler, respectively. The results of the paper are summarized in Section~\ref{sec:X}. The paper has four appendices containing some details of the derivations. 

We use the Gaussian system of units throughout this paper.

%
\section{Energy change of the beam due to coherent radiation}\label{sec:II}
%

%
\subsection{Derivation of the energy change using retarded potentials}\label{sec:IIA}
%

We begin from the equation that describes the rate of change of energy $\mathcal E$ of a  point charge $e$ moving in electric field $\vec E(\vec r,t)$ with velocity $\vec v$,
    \begin{align}\label{eq:1}
    \frac{d \mathcal E}{dt}
    =
    e\vec v\cdot\vec E
    .
    \end{align}
Expressing the electric field through the scalar potential $\phi(\vec r,t)$ and the vector potential $\vec A(\vec r,t)$, $\vec E = -\nabla \phi-c^{-1}\p_t\vec A$, it is easy to cast~\eqref{eq:1} into the following form,
    \begin{align}
    \label{eq:2}
    \frac{d (\mathcal E + e\phi)}{dt}
    =
    e\frac{\p \phi}{\p t}
    -
    e\vec{\beta}\cdot\frac{\p\vec{A}}{\p t}
    ,
    \end{align}
where $\vec \beta = \vec v/c$ and $c$ is the speed of light. In Eq.~\eqref{eq:2} the full time derivative $d\phi/dt = \p_t\phi+\vec v\cdot\nabla\phi$ is taken along the particle orbit and gives the rate of change of $\phi$ as seen by the moving charge.

Eqs.~\eqref{eq:1} and~\eqref{eq:2} are valid for a point charge. To apply them to a beam of charged particles, we represent the latter as a cold fluid that is characterized by the charge density $en(\vec r,t)$ and the fluid velocity $\vec v(\vec r,t)$, where $n(\vec r,t)$ is the particle density. The current density in the beam is $\vec j(\vec r,t) = e n(\vec r,t)\vec v(\vec r,t)$. Note that in this description of the beam we neglect the effects of the beam emittance and  energy spread and at a given time $t$ associate a unique value of the velocity $\vec v$ with each location $\vec r$ within the bunch. With this new understanding of the velocity field $\vec{v}(\vec{r},t)$, Eq.~\eqref{eq:2} can be written as
    \begin{align}\label{eq:3}
    \frac{d (\mathcal E + e\phi)}{dt}
    &=
    e\frac{\p (\phi-\vec{\beta}\cdot\vec{A})}{\p t}
    +
    e\vec{A}\cdot\frac{\p\vec{\beta}}{\p t}
    =
    e\frac{\p V}{\p t}
    +
    e\vec{A}\cdot\frac{\p\vec{\beta}}{\p t}
    ,
    \end{align}
where $V(\vec r,t) = \phi(\vec r,t)-\vec{\beta}(\vec r,t)\cdot\vec{A}(\vec r,t)$. The function $V$ was first introduced into the calculation of CSR wakefields in Ref.~\cite{derbenev95rss}.

We will limit our consideration to the cases where the velocity $\vec v$ at a given location $\vec r$ does not depend on time $t$, $\vec v = \vec v(\vec r)$, which is a good approximation for relativistic beams with a small angular spread when all the particles at a given location are approximately moving in one direction---the direction of the tangent vector to the trajectory of the reference particle. In this case, the last term on the right-hand side of~\eqref{eq:3} can be neglected, and the rate of change of $\mathcal E + e\phi$ is given by the partial time derivative of $V$.
    
In free space, far from metal boundaries, $\phi$ and $\vec A$ are expressed in terms of $n$ and $\vec v$ through the retarded potentials~\cite{landau_lifshitz_ctf},
    \begin{align}\label{eq:4}
    \phi(\vec r,t)
    &=
    \frac{e}{c}
    \int
    \frac{d^3r'}{\tau}
    n(\vec r',t-\tau)
    ,
    \nonumber\\
    \vec A(\vec r,t)
    &=
    \frac{e}{c}
    \int
    \frac{d^3r'}{\tau}
    \vec \beta(\vec r')
    n(\vec r',t-\tau)
    ,
    \end{align}
where $\tau=\tau(\vec r,\vec r') = |\vec r-\vec r'|/c$. Correspondingly, for function $V$ one finds
    \begin{align}\label{eq:5}
    V(\vec r,t)
    =
    \frac{e^2}{c}
    \int
    \frac{d^3r'}{\tau}
    (1-\vec \beta\cdot\vec\beta')
    n(\vec r',t-\tau)
    ,
    \end{align}
where $\vec \beta = \vec\beta(\vec r)$ and $\vec \beta' = \vec\beta(\vec r')$. Note that the integrand in this expression has a singularity when $\vec r' \to \vec r$ because at this point $\tau=0$. This singularity however is integrable in three (and two) dimensions, and the function $V$ is finite. 

Considerable simplifications can be achieved if one chooses a line charge model for the beam. In this model, all the particles in the beam are moving on the same orbit $\vec r_0(s)$ parametrized by the arc length $s$ measured along it. The vector $\vec\beta(s)$ is directed along the tangent vector to the orbit. The distribution function in 1D is denoted by $\lambda(s,t)$; it gives the number of particles per unit $s$. Eq.~\eqref{eq:5} is now written as a one-dimensional integral,
    \begin{align}\label{eq:6}
    V(s,t)
    =
    \frac{e^2}{c}
    \int_{-\infty}^\infty
    \frac{ds'}{\tau}
    (1-\vec \beta\cdot\vec\beta')
    \lambda(s',t-\tau)
    ,
    \end{align}
where $\tau(s,s')=|\vec r_0(s)-\vec r_0(s')|/c$ is a function of $s$ and $s'$, and $\vec \beta = \vec\beta(s)$, $\vec \beta' = \vec\beta(s')$ are defined on the orbit. Note that, in general case, the singularity of the integrand in the limit $s' \to s$ makes the integral (logarithmically) divergent, unless the particles are moving with the speed of light\footnote{A different approach to eliminate the singularity without the assumption $v=c$ was used in~\cite{saldin97sy,saldin98sy}: the term responsible for the singularity was called the space charge effect; it was isolated and discarded as not relevant to the radiation wakefield.}. In this latter case,  $|\vec \beta|=|\vec \beta'|=1$ and in the limit $s'\to s$ the term $1-\vec \beta\cdot\vec\beta'$ tends to zero canceling the vanishing $\tau$. Hence, in what follows we assume $|\vec \beta|=1$, which means that all the particles in the bunch are moving with the speed of light. In accordance with this assumption, the distribution function $\lambda$ is transported along the orbit without changing its shape and can be written as a function $\lambda(s-ct)$ of one argument $s-ct$. With this distribution function, Eq.~\eqref{eq:6} can be rewritten,
    \begin{align}\label{eq:7}
    V(s,t)
    =
    \frac{e^2}{c}
    \int_{-\infty}^\infty
    \frac{ds'}{\tau}
    (1-\vec \beta\cdot\vec\beta')
    \lambda(s'-c(t-\tau))
    .
    \end{align}
Using Eq.~\eqref{eq:3} (in which we agreed to neglect the last term), we obtain
    \begin{align}\label{eq:8}
    \frac {d\cal E}{dt} 
    +
    e
    \frac{d\phi}{dt}
    = 
    \frac{\p V}{\p t}
    &=
    -{e^2}
    \int_{-\infty}^\infty
    \frac{ds'}{\tau}
    (1-\vec \beta\cdot\vec\beta')
    \lambda'(s'-c(t-\tau))
    ,
    \end{align}
where $\lambda'$ denotes the derivative of function $\lambda$ with respect to its argument.

Our general setup for a class of problems considered in this paper consists of a region of space occupied by time independent magnetic field (a bending magnet or a sequence of magnets, an undulator, etc.). Before entering this region, and after exiting it, the beam travels along straight lines. Our goal will be to calculate the energy loss $\Delta {\cal E}(z)$ (where $z$ is the longitudinal coordinate inside the beam) of different slices of the beam after it propagates sufficiently far enough from the exit ($t\to\infty$), so that its electromagnetic field returns to a steady state (the same state the beam had before entering the region, at $t\to-\infty$). In this calculation, we will assume that the potential at each particle of the bunch in the final state is the same as initial, $\Delta\phi = \phi(t\to\infty)-\phi(t\to-\infty)=0$. This assumption is justified if the bunch is not focused transversely or compressed longitudinally relative to its initial state after it passes through the region of magnetic field. The effects of transverse focusing  on $\phi$ in round pipes were studied in Ref.~\cite{bane02c}; in principle, they can be added to our formalism, but they are not a subject of this work.

Taking into account the condition $\Delta \phi = 0$, integration of~\eqref{eq:8} over time from minus to plus infinity gives the energy change $\Delta\mathcal{E}$ from the initial to the final state. Because of the full derivative $d{\cal E}/dt$, the integration has to be carried out along the particle trajectory $s=z+ct$, where $z$ is an integral of motion and is equal to the coordinate $s$ of a slice in the beam at $t=0$. Replacing $s$ by $z+ct$ on the right-hand side of~\eqref{eq:8}, we integrate it over time,
    \begin{align}\label{eq:9}
    \Delta {\cal E}(z)
    &=
    -
    {e^2}
    \int_{-\infty}^{\infty}
    dt
    \int_{-\infty}^{\infty}
    \frac{ds'}{\tau(z+ct,s')}
    (1-\vec \beta(z+ct)\cdot\vec\beta(s'))
    \lambda'(s'-ct+c\tau(z+ct,s'))
    \nonumber\\
    &=
    -
    \frac{e^2}{c}
    \int_{-\infty}^{\infty}
    ds
    \int_{-\infty}^{\infty}
    \frac{ds'}{\tau(s,s')}
    (1-\vec \beta(s)\cdot\vec\beta(s'))
    \lambda'(s'-s+z+c\tau(s,s'))
    .
    \end{align}
In the last integral we replaced the integration over time by integration over $s$. Formula~\eqref{eq:9} gives the total integrated energy change at coordinate $z$ in the bunch. 

We will also consider in the paper the two cases when the asymptotic trajectories at $t\to\pm \infty$ are not straight lines: these are the case of a circular motion \cite{derbenev95rss,murphy97kg} and an infinitely long wiggler \cite{saldin98sy,wu03rs}. These two models represent a long enough region of the magnetic field, such that the transient effects due to the entrance to and exit from it can be neglected. In these two cases the relevant quantity is the energy loss \emph{per unit length} (averaged over the wiggler period in the case of the wiggler). For circular motion the integration over $s$ in~\eqref{eq:9} is omitted and the formula gives an energy loss per unit length. For an infinitely long wiggler the integration over $s$ is replaced by averaging over one period of the wiggler.

As was first pointed out in~\cite{derbenev95rss}, and also in subsequent studies, for short bunches, the main contribution to the integral~\eqref{eq:7} comes from the particles behind the observation point, that is $s'<s$. While Eqs.~\eqref{eq:6}--\eqref{eq:9} are valid for arbitrary bunch length, in this paper, following~\cite{derbenev95rss}, we will limit our analysis to such short bunches and replace the infinite upper limit in the integral over $s'$ by $s$:
    \begin{align}\label{eq:10}
    \Delta {\cal E}(z)
    &=
    -
    \frac{e^2}{c}
    \int_{-\infty}^{\infty}
    ds
    \int_{-\infty}^{s}
    \frac{ds'}{\tau(s,s')}
    (1-\vec \beta(s)\cdot\vec\beta(s'))
    \lambda'(s'-s+z+c\tau(s,s'))
    .
    \end{align}
In many subsequent equations of this section this assumption can be easily omitted and, if needed, the original form~\eqref{eq:9} used instead of~\eqref{eq:10}. 

%
\subsection{CSR wake and impedance}\label{sec:IIB}
%

Instead of working with function $\Delta {\cal E}(z)$ it is more convenient to introduce the radiation longitudinal wake $w(z)$ and impedance $Z(k)$. The wake $w(z)$ of a point charge is defined by the following relation (see, e.g.,~\cite{chao93})
    \begin{align}\label{eq:11}
    \Delta {\cal E}(z) 
    =
    -
    e^2
    \int_{-\infty}^\infty
    \lambda(z')w(z-z')
    dz'
    =
    -
    e^2
    \int_{-\infty}^\infty
    \lambda(z-\zeta)w(\zeta)
    d\zeta
    ,
    \end{align}
where $\zeta=z-z'$. In this formula we do not assume that the wake is localized in front of or behind the particle---an assumption often used in the standard wakefield theory. The sign of the wake $w$ is chosen so that a positive $w$ corresponds to energy loss. The longitudinal impedance is defined by
    \begin{align}\label{eq:12}
    Z(k)
    =
    \frac{1}{c}
    \int_{-\infty}^\infty
    dz w(z)
    e^{-ik z}
    .
    \end{align}
Following Ref.~\cite{stupakov02h}, we use here $e^{-ikz}$ because the coordinate $z$ is measured in the direction of motion (in contrast to the classical wakes where $z$ is often measured in the opposite direction). Combining Eqs.~\eqref{eq:11} and~\eqref{eq:12} we obtain
    \begin{align}\label{eq:13}
    \Delta {\cal E}(z) 
    &=
    -
    e^2
    c
    \int_{-\infty}^\infty
    dk Z(k)
    \hat\lambda(k)
    e^{ik z}
    =
    -
    2e^2
    c
    \Re
    \int_{0}^\infty
    dk Z(k)
    \hat\lambda(k)
    e^{ik z}
    ,
    \end{align}
where
    \begin{align}\label{eq:14}
    \hat\lambda(k)
    =
    \frac{1}{2\pi}
    \int_{-\infty}^\infty
    dz'
    e^{-ikz'}
    \lambda(z')
    \end{align}
is the Fourier transform of the distribution function. Making the inverse Fourier transform of~\eqref{eq:13} we express $Z$ through $\Delta {\mathcal E}$ 
    \begin{align}\label{eq:15}
    Z(k)
    =
    -
    \frac{1}{2\pi e^2c\hat\lambda(k)}
    \int_{-\infty}^\infty
    dz
    \Delta {\mathcal E}(z) 
    e^{-ik z}
    .
    \end{align}
Substituting~\eqref{eq:10} into this equation, changing the integration variable from $z$ to $s=z+ct$ and  carrying out the integration over $t$ gives the following result:
    \begin{align}\label{eq:16}
    Z(k)
    =
    \frac{ik}{c^2}
    \int_{-\infty}^{\infty}
    ds
    \int_{-\infty}^{s}
    \frac{ds'}{\tau(s,s')}
    (1-\vec \beta(s)\cdot\vec\beta(s'))
    e^{ik(c\tau(s,s')-s+s')}
    .
    \end{align}
We see that the distribution function $\hat \lambda$ is disappears from the definition of $Z$, as expected.  Being a Fourier transform of the real function $w$ (see Eq.~\eqref{eq:12}) the impedance has a property $Z(-k) = Z^*(k)$.

A useful formula for the total energy loss $U$ of the bunch due to radiation can be obtained from Eq.~\eqref{eq:13},
    \begin{align}\label{eq:17}
    U
    \equiv
    -
    \int_{-\infty}^\infty
    dz\,
    \Delta {\cal E}(z)
    \lambda(z)
    =
    4\pi
    e^2
    c
    \int_{0}^\infty
    dk Z(k)
    |\hat\lambda(k)|^2
    .
    \end{align}

%
\section{Divergence of free-space impedance and necessity of shielding}\label{sec:III}
%

Eq.~\eqref{eq:16} gives a general formula for calculation of the impedance for arbitrary beam trajectory. As we show in Section~\ref{sec:V}, it can  easily  be applied to an infinitely long wiggler and a circular orbit (in the latter case the integration over $s$ in~\eqref{eq:16} is dropped), and reproduces the known results. Unfortunately, the integral over $s$ diverges for the  trajectories that begin and end as straight lines. This statement will be proved in Appendix~\ref{app:B} for the case of a bending magnet; it also follows from the expression for the CSR wake derived in Ref.~\cite{stupakov02e}. There is a simple physical mechanism behind this divergence: it is due to the edge radiation~\cite{bosch98} of the beam at the entrance and the exit from the magnet. Indeed, the spectral energy at a given frequency $\omega$ of the edge radiation of a relativistic particle is proportional to $\ln\gamma$ and tends to infinity when $\gamma\to\infty$. At the same time, the spectral energy loss of the beam due to radiation at this frequency is proportional to the real part of $Z(\omega/c)$, see Eq.~\eqref{eq:17}; this explains the divergence of $Z$ in the limit $\gamma\to\infty$. In many practical cases, the circumstance that makes the energy of the edge radiation finite is the presence of metal walls of the vacuum chamber surrounding the orbit, or shielding.

\begin{figure}[htb]
    \centering
    \includegraphics[width=0.6\textwidth, trim=0mm 0mm 0mm 0mm, clip]{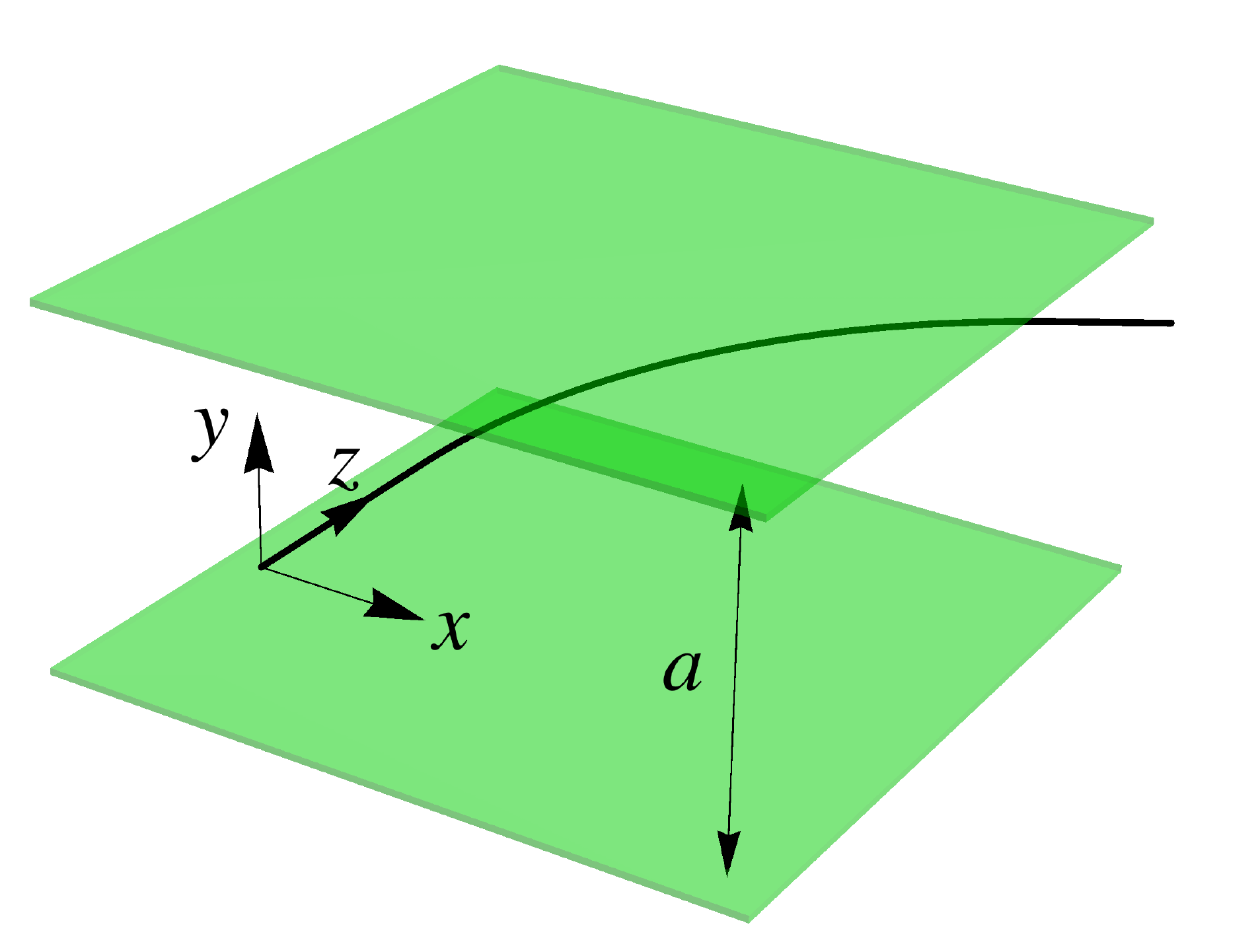}
    \caption{Parallel plates located at $y=\pm \frac{1}{2}a$ and a plane orbit 
    in the mid-plane shown by black. The coordinate system is chosen so that the orbit lies in the 
    $xz$ plane
    and the $z$-axis is directed along the tangent vector to the trajectory at $x=z=0$.}
    \label{fig:1}
    \end{figure}

The simplest model that takes into account the shielding and at the same time allows for analytical results consists of two parallel perfectly conducting plates with the orbit located in the middle as shown in Fig.~\ref{fig:1}. 
We will assume that the plates are located at $y=\pm \frac{1}{2}a$ with $a$ being the full gap between the plates.

The derivation of $\Delta {\cal E}(z)$ and $Z(k)$ from the previous section can be easily generalized to include the boundary conditions at the metal plates. These conditions require zero tangential electric field on the surface of the plates and can be satisfied by introducing image charges and currents to the system~\cite{murphy97kg}. With account of these image charges and currents, Eqs.~\eqref{eq:4} are replaced by
    \begin{align}\label{eq:18}
    \phi(\vec r,t)
    &=
    \frac{e}{c}
    {\sum_{m=-\infty}^\infty}
    \int
    \frac{d^3r'}{\tau}
    n_m(\vec r',t-\tau)
    ,
    \nonumber\\
    \vec A(\vec r,t)
    &=
    \frac{e}{c}
    {\sum_{m=-\infty}^\infty}
    \int
    \frac{d^3r'}{\tau}
    \vec \beta_m (\vec r')
    n_m(\vec r',t-\tau)
    .
    \end{align}
Here index $m$ marks the images with the charge density
    \begin{align}\label{eq:19}
    en_m(\vec r,t)
    =
    (-1)^m
    en_0(\vec r-m a \hat{\vec y},t)
    ,
    \end{align}
where $\hat{\vec y}$ is the unit vector in $y$ direction and $n_0(\vec r,t)$ is the density distribution of the ``real'' beam. The normalized velocity of the $m$-th image is
    \begin{align}\label{eq:20}
    \beta_{m,x} (\vec r)
    =
    \beta_{0,x} (\vec r-m a \hat{\vec y})
    ,\qquad
    \beta_{m,y} (\vec r)
    =
    (-1)^m
    \beta_{0,y} (\vec r-m a \hat{\vec y})
    ,\qquad
    \beta_{m,z} (\vec r)
    =
    \beta_{0,z} (\vec r-m a \hat{\vec y})
    .
    \end{align}
In what follows, we consider plane orbits lying in the $y=0$ plane. For such orbits  $\beta_{0,y}=0$ and Eqs.~\eqref{eq:20} are simplified,
    \begin{align}\label{eq:21}
    \vec\beta_{m} (\vec r)
    =
    \vec\beta_{0} (\vec r-m a \hat{\vec y})
    .
    \end{align}

Substituting~\eqref{eq:19} and~\eqref{eq:21} into~\eqref{eq:18} and changing the integration variable $\vec r'-m a \hat{\vec y}\to \vec r'$ we obtain
    \begin{align}\label{eq:22}
    \phi(\vec r,t)
    &=
    \frac{e}{c}
    {\sum_{m=-\infty}^\infty}
    (-1)^m
    \int
    \frac{d^3r'}{\tau_m}
    n_0(\vec r',t-\tau_m)
    ,
    \nonumber\\
    \vec A(\vec r,t)
    &=
    \frac{e}{c}
    {\sum_{m=-\infty}^\infty}
    (-1)^m
    \int
    \frac{d^3r'}{\tau_m}
    \vec \beta_0(\vec r') 
    n_0(\vec r',t-\tau_m)
    ,
    \end{align}
where $c\tau_m(\vec r,\vec r') = |\vec r-\vec r'+m a \hat{\vec y}|$. Replacing Eqs.~\eqref{eq:4} by Eqs.~\eqref{eq:22} and repeating the derivation of the  impedance $Z(k)$ from the previous Section with the new expressions for the potentials, we obtain a generalization of Eq.~\eqref{eq:16} that includes the effect of shielding by parallel plates:
    \begin{align}\label{eq:23}
    Z(k)
    =
    \frac{ik}{c^2}
    \int_{-\infty}^{\infty}
    ds
    \int_{-\infty}^{s}
    ds'
    {\sum_{m=-\infty}^\infty}
    (-1)^m
    \frac{1-\vec \beta(s)\cdot\vec\beta(s')}{\tau_m(s,s')}
    e^{ik(c\tau_m(s,s')-s+s')}
    ,
    \end{align}
where $c\tau_m(s,s')=\sqrt{|\vec r_0(s)-\vec r_0(s')|^2+m^2a^2}$ and we replaced $\vec \beta_0$ by the original notation $\vec \beta$. Our previous result~\eqref{eq:16} of the impedance in free space is contained in this formula as a summand with $m=0$.

While Eq.~\eqref{eq:23} looks like a viable starting point for practical calculations, there are two difficulties associated with it. The first one is that the summation over $m$ cannot be interchanged with the integration, because, as mentioned above, the integration in the term $m=0$ diverges in the case of straight orbits. The second difficulty is due to a slow convergence of the sum over $m$ with the subsequent summands changing sign. These two difficulties can be overcome through a transformation in which the summation over $m$ is carried out. This transformation is described in Appendix~\ref{app:1}; it replaces~\eqref{eq:23} by the following formula,
    \begin{align}\label{eq:24}
    Z(k)
    &=
    -
    \frac{2\pi k}{ac}
    \sum_{p=0}^\infty
    \int_{-\infty}^{\infty}
    ds
    \int_{-\infty}^{s}
    ds'
    H_0^{(1)}
    \left(
    ck\tau(s,s')
    \sqrt{1-(2p+1)^2\pi^2/k^2a^2}
    \right)
    \nonumber\\
    &\times
    (1-\vec \beta(s)\cdot\vec\beta(s'))
    e^{-ik(s-s')}
    ,
    \end{align}
where $H_0^{(1)}$ is the Hankel function of the first kind.

We now make too simplifying assumptions. We first assume that
    \begin{align}\label{eq:25}
    ka
    \gg
    1
    ,
    \end{align}
that is the reduced wavelength $\lambdabar = \lambda/2\pi$ that can be associated with the bunch length is much smaller than the gap $a$ between the plates. With the sum over $p$ rapidly converging, this allows us to treat $\pi(2p+1)/ka$ is a small parameter. Second, we assume that $ck\tau$ in the argument of the Hankel function in the region of integration that makes a dominant contribution to the integral is much greater than one. This typically means that
    \begin{align}\label{eq:26}
    k\rho
    \gg
    1
    ,
    \end{align}
or the reduced wavelength is much smaller than the characteristic bending radius in the system. With these two assumptions we Taylor expand the square root in the argument of $H_0^{(1)}$ in~\eqref{eq:24} and use the asymptotic expansion for the Hankel function in the limit of large argument,
    \begin{align}\label{eq:27}
    H_0^{(1)}(z)
    \approx
    (1-i)
    \sqrt{\frac{1}{\pi z}}
    e^{iz}
    .
    \end{align} 
We then obtain
    \begin{align}\label{eq:28}
    Z(k)
    &\approx
    (i-1)
    \frac{2\sqrt{\pi k}}{ac}
    \sum_{p=0}^\infty
    \int_{-\infty}^{\infty}
    ds
    \int_{-\infty}^{s}
    ds'
    \frac{1-\vec \beta(s)\cdot\vec\beta(s')}{\sqrt{|\vec r_0(s)-\vec r_0(s')|}}
    \exp
    \left(
    -i
    c\tau(s,s')    
    \frac{(2p+1)^2\pi^2}{2ka^2}
    \right)
    \nonumber\\
    &\times
    \exp\left[
    ik(
    c\tau(s,s')    
    -s+s')
    \right]
    .
    \end{align}

One more approximation can be made if we assume that the trajectory is at a small angle with a straight line.  We then choose a Cartesian coordinate system with coordinate $z$ directed along this line and approximate $c\tau(s,s')=|\vec r_0(s)-\vec r_0(s')|\approx z-z'$ in~\eqref{eq:28}:
    \begin{align}\label{eq:29}
    Z(k)
    &=
    (i-1)
    \frac{2\sqrt{\pi k}}{ac}
    \sum_{p=0}^\infty
    \int_{-\infty}^{\infty}
    dz
    \int_{-\infty}^{z}
    dz'
    \frac{1-\vec \beta(z)\cdot\vec\beta(z')}{\sqrt{z-z'}}
    \exp
    \left(
    -i(z-z')
    \frac{(2p+1)^2\pi^2}{2ka^2}
    \right)
    \nonumber\\
    &\times
    \exp\left[
    ik(
    c\tau(z,z')    
    -s(z)+s(z'))
    \right]
    .
    \end{align}
Note that replacing $c\tau$ by its approximation $z-z'$ in two places of the integrand in~\eqref{eq:28} we do not do this in the last exponent. The reason for that is that, as we mentioned above, $ck\tau$ is a large number, and even small corrections to it can lead to a large phase error in the last exponential function. A more accurate approximation for this term will be used in subsequent sections.

%
\section{Computer code CSRZ}\label{sec:IV}
%

To verify the validity of approximations that were made in the derivation of the radiation impedance, in the following sections of the paper we make a comparison of our analytical results with a computer code CSRZ that uses a numerical algorithm to find electromagnetic field of a relativistic bunch and calculate the longitudinal wake and impedance. The details of the algorithm implemented in the code can be found in Ref.~\cite{Zhou2012}. Here we give its brief description.

The code solves the parabolic equation~\cite{stupakov03k, agoh04,agoh04y} in the frequency domain in a curvilinear coordinate system $x,y,s$,
    \begin{equation}
    \frac{\partial \vec{E}_\perp}{\partial s}
    =
    \frac{i}{2k}
    \left(\nabla_\perp^2\vec{E}_\perp
    -
    {4\pi e}\nabla_\perp\, n 
    + 
    \frac{2k^2x}{\rho (s)}\vec{E}_\perp 
    \right)
    ,
    \label{eq:30}
    \end{equation}
where $\vec{E}_\perp=(E_x, E_y)$ is the transverse electric field and $k=\omega/c$ is the wavenumber. The boundary conditions for the field correspond to a metal surface of a rectangular cross section with a given aspect ratio $b/a$ (where $a$ is the size of the rectangle along $y$ and $b$ is along $x$). The beam has transverse charge distribution $en(x,y)$ that is independent of $s$. In calculations presented in this paper we used a bi-Gaussian transverse distribution with the rms sizes of a few tens of microns in the vertical and a few hundreds of microns in the horizontal directions. The radius of curvature of the reference orbit $\rho(s)$ is allowed to arbitrary vary along $s$.  Specifying different functions $\rho(s)$ enables the code to simulate a broad range of practical devices, such as a single bending magnet, a series of bending magnets connected by straight chambers, or even an undulator or a wiggler. 

With paraxial approximation~\cite{agoh04y}, the longitudinal electric field is found to be
    \begin{equation}
    E_s=\frac{i}{k} 
    \left(\nabla_\perp \cdot \vec{E}_\perp 
    - 
    \frac{4\pi}{c}j_s
    \right)
    ,
    \label{eq:31}
    \end{equation}
where $j_s=enc$ is the current density. Then longitudinal radiation impedance is calculated by directly integrating $E_s$ over $s$
    \begin{equation}
    Z_\parallel(k)
    =
    -\frac{1}{Q}\int_{-\infty}^\infty E_s(x_c,y_c,s) ds
    ,
    \label{eq:32}
    \end{equation}
where $(x_c,y_c)$ denotes the center of the beam in the transverse plane and $Q$ is the charge of the beam.

While the code calculates the impedance assuming a metallic vacuum chamber of rectangular cross section, our analytical theory deals with two parallel metal plates. To be able to do a comparison between the two approaches we set the vertical dimension of the vacuum chamber in the code equal to the gap between the plates. At the same time, to minimize the effect of the vertical walls of the chamber, we choose a large aspect ratio $b/a$. This positions the vertical walls far from the beam orbit and suppresses their effect on the impedance. Experimenting with various aspect ratios, we found that a good agreement with the parallel plates model can be achieved if the aspect ratio $b/a\gtrsim 3$. Below we indicate in the text the aspect ratio used in each particular simulation.

%
\section{Reproducing known results}\label{sec:V}
%

In this section we will show how some of the known analytical results for the CSR impedance can be easily obtained from the general formalism developed in the Section~\ref{sec:III}.

%
\subsection{Circular orbit}\label{sec:VA}
%

\begin{figure}[htb]
    \centering
    \includegraphics[width=0.6\textwidth, trim=0mm 0mm 0mm 0mm, clip]{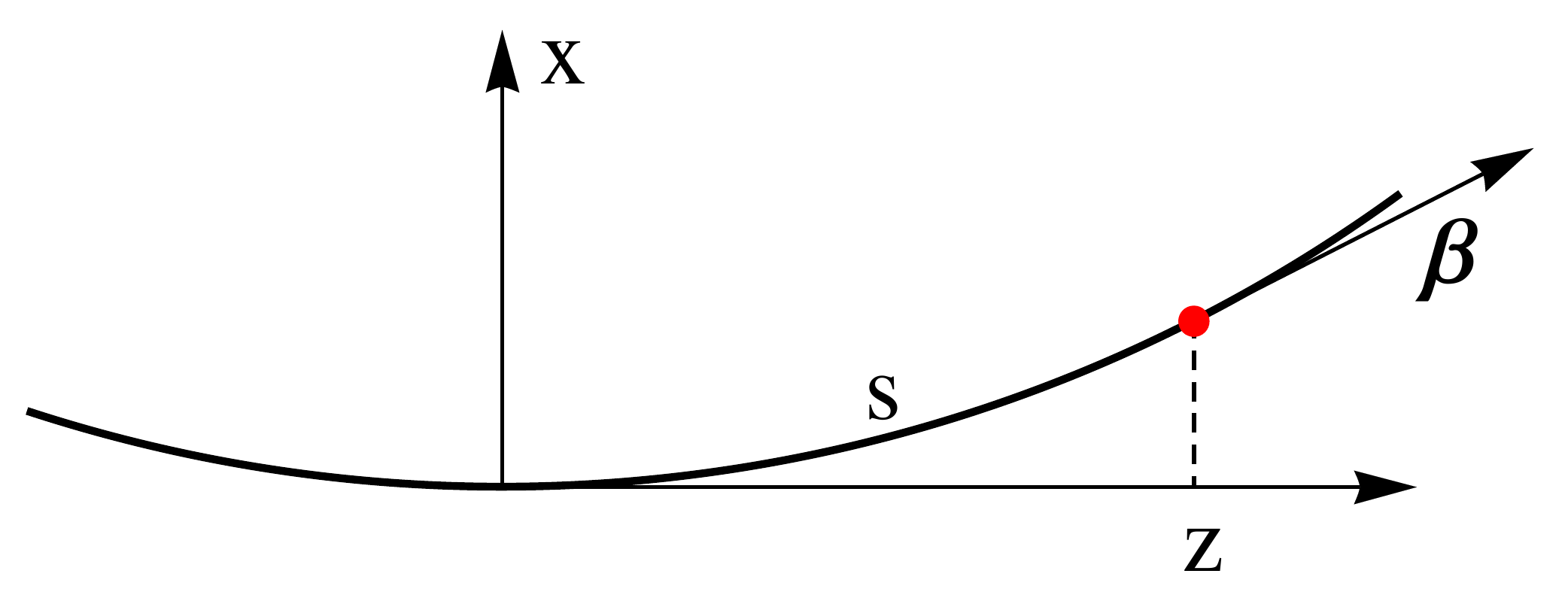}
    \caption{A part of a circular orbit with coordinate system.}
    \label{fig:2}
    \end{figure}

We first consider a circular orbit of radius $\rho$ and calculate the CSR impedance $Z$ per unit length. The coordinate system and the orbit are shown in Fig.~\ref{fig:2}.  

For large values of the wavenumber, $k\gg 1/\rho$, the dominant contribution to the integrals~\eqref{eq:16} comes from distances much smaller then $\rho$, and we can use approximate formulas for the orbit:
    \begin{align}
    \label{eq:33}
    x_0(z)
    =
    \frac{1}{2\rho}
    z^2
    ,
    \qquad
    y_0(z)
    =
    0
    .
    \end{align}	                                                                                                                                                                                                                                                                                                                                                                                                                                                                                                                                                                                                                                                                                                                                                                                                                                                                                                                                 
Within the same approximation, vector $\vec \beta$ is given by
    \begin{align}
    \label{eq:34}
    \vec{\beta}_\perp(z)
    &=
    \hat{\vec{x}} 
    \frac{z}{\rho}
    ,\qquad
    \beta_z
    =
    1
    -
    \frac{1}{2\rho^2}
    z^2
    .
    \end{align}	                                                                                                                                                                                                                                                                                                                                                                                                                                                                                                                                                                                                                                                                                                                                                                                                                                                                                                                                 
We now express $\tau$ 
    \begin{align}\label{eq:35}
    c\tau
    =
    \sqrt{
    (z-z')^2
    +
    (x_0(z)-x_0(z'))^2
    }
    \approx
    |z-z'|
    +
    \frac{1}{8\rho^2|z-z'|}
    (z^2-z'^2)^2
    ,
    \end{align}
and the factor $1-\vec \beta\cdot\vec\beta'$, 
    \begin{align}\label{eq:36}
    1-\vec \beta\cdot\vec\beta'
    &=
    1
    -
    \left(
    1
    -
    \frac{1}{2\rho^2}
    z^2
    \right)
    \left(
    1
    -
    \frac{1}{2\rho^2}
    z'^2
    \right)
    -
    \frac{1}{\rho^2}
    zz'
    \approx
    \frac{1}{2\rho^2}
    \left(
    z
    -
    z'
    \right)^2
    ,
    \end{align}
as functions of $z$ and $z'$. Finally, integrating the  relation for the arc length
    \begin{align}\label{eq:37}
    \frac{ds}{dz}
    =
    \sqrt{
    1
    +
    \left(\frac{dx_0}{dz}\right)^2
    }
    \approx
    1
    +
    \frac{1}{2\rho^3}
    z^2
    ,
    \end{align}
we obtain
    \begin{align}\label{eq:38}
    s(z)
    \approx
    z
    \left(1+ \frac{1}{6\rho^2}z^2\right)
    .
    \end{align}
 
For unshielded CSR impedance (corresponding to the gap between the conducting plates $a\to\infty$) we can use Eq.~\eqref{eq:16}, in which, as discussed in Section~\ref{sec:IIA}, we drop the integration over $s$; this gives the impedance per unit length of the trajectory. We also replace integration over $s'$ by integration over $z'$, $ds'\approx dz'$ and take into account that $z'<z$ which means that the wake acting on a given particle in the bunch is determined by the particles behind it. For $\tau$ in the denominator of~\eqref{eq:16} we use $\tau \approx (z-z')/c$, while more accurate expressions~\eqref{eq:35} and~\eqref{eq:38} are substituted into the exponent. The result is:
    \begin{align}\label{eq:39}
    Z(k)
    &=
    \frac{1}{2}
    \frac{ik}{c\rho^2}
    \int_{-\infty}^z
    {dz'}
    (z-z')
    \exp\left(
    -
    ik
    \frac{1}{24\rho^2}(z-z)'^3
    \right)
    =
    \frac{1}{3^{1/3}}
    (i+\sqrt{3})
    \Gamma\left( \frac{2}{3} \right)
    \frac{k^{1/3}}{c\rho^{2/3}}
    ,
    \end{align}
where $\Gamma$ is the gamma-function. Note that the main contribution to the integral comes from the distance $z-z'\sim \ell_\|$, with $\ell_\| = (24\rho^2/k)^{1/3}$; this distance is interpreted as the \emph{formation length} of the radiation with the wavelength $2\pi/k$. Eq.~\eqref{eq:39} fully agrees with the wakefield first derived in~\cite{derbenev95rss} (and our derivation to some extent repeats the derivation in that paper).

It is also very easy to derive the shielded CSR impedance of a circular orbit between parallel conducting plates, and reproduce the result of Ref.~\cite{murphy97kg}. For this we use Eq.~\eqref{eq:29} again dropping the integration over $z$ to obtain the impedance per unit length,
    \begin{align}\label{eq:40}
    Z(k)
    &=
    (i-1)
    \frac{\sqrt{\pi k}}{ac\rho^2}
    \sum_{p=0}^\infty
    \int_{-\infty}^{z}
    dz'
    \left(
    z
    -
    z'
    \right)^{3/2}
    \exp
    \left(
    -i(z-z')
    \frac{(2p+1)^2\pi^2}{2ka^2}
    -i
    \frac{k}{24\rho^2}(z-z)'^3
    \right)
    .
    \end{align}
To improve the convergence of the integral we change the integration variable from $z'$ to $t$ with $t=e^{i\pi/6}(k/24\rho^2)^{1/3}(z-z')$ which corresponds to the rotation of the integration path in the complex plane of the variable $z-z'$. As a result we arrive at the expression for the impedance in the form first obtained in Ref.~\cite{murphy97kg},
    \begin{align}\label{eq:41}
    Z(k)
    &=
    \frac{4\sqrt{2\pi}}{c}
    3^{2/3}
    e^{i\pi/6} 
    \frac{k^{1/3}}{\alpha\rho^{2/3}}
    {\sum_{p=0}^\infty}
    \int_0^{\infty}
    t^{3/2} dt
    \exp
    \left(
    {-t^3-t
    \frac{(2p+1)^2\pi^2}{2\alpha^2}}
    \right)
    ,
    \end{align}
where
    \begin{align}\label{eq:42}
    \alpha=\frac{e^{-i\pi/6}k^{2/3}a}{2^{1/2}3^{1/6}\rho^{1/3}}
    \,.
    \end{align}
We see that the impedance, apart from a general scaling factor, depends on one dimensionless variable $k^{2/3}a/\rho^{1/3}$ which can be interpreted as a ratio of $a$ to the \emph{transverse coherence (or formaion) size} of the radiation $\ell_\perp\sim\rho^{1/3}/k^{2/3}$. Analysis shows that in the limit $a\gg\ell_\perp$ the shielded impedance~\eqref{eq:41} approaches the unshielded result~\eqref{eq:39}. In the opposite limit, $a\ll\ell_\perp$, the shielded impedance becomes much smaller that~\eqref{eq:39}.

%
\subsection{Infinitely long wiggler in free space}\label{sec:VB}
%

CSR wake of an infinitely long wiggler in free space was first calculated in Ref.~\cite{saldin97sy}. The complicated general analytical expressions derived in that paper were somewhat simplified in Ref.~\cite{stupakov02e} in the limit $v=c$ and assuming the wiggler parameter $K\gg 1$. It was then used in the study of the beam instability in damping rings in~\cite{wu03rsh}. We will now show how the result of~\cite{stupakov02e} can be straightforwardly obtained from the method developed in this work. The derivation below is much simpler than the approach used in Ref.~\cite{wu03rsh}.

Consider a long plane wiggler that is characterized by the wiggler parameter $K\gg 1$ and period $\lambda_w$. In this analysis we neglect the contribution to the impedance from the transient regions at the entrance to and the exit from the wiggler. The trajectory of a relativistic particle with the Lorentz factor $\gamma$ in such  a wiggler is given by
    \begin{align}\label{eq:43}
    x_0(z)=-\frac{\theta_0}{k_w}
    \cos k_wz,
    \qquad 
    y_0(z)=0,
    \end{align}
with the velocity 
    \begin{align}
    \label{eq:44}
    \vec{\beta}_\perp(z)
    &=
    \theta_0
    \hat{\vec{x}}\sin k_wz
    ,\qquad
    \beta_z
    =
    1
    -
    \frac{1}{2}
    \theta_0^2
    \sin^2 k_wz
    ,
    \end{align}	                                                                                                                                                                                                                                                                                                                                                                                                                                                                                                                                                                                                                                                                                                                                                                                                                                                                                                                                 
where $k_w=2\pi/\lambda_w$,  $\theta_0 = K/\gamma$ and we assume that $\theta_0\ll 1$. Note, that we take the limit $\gamma\to\infty$ after we have introduced $\theta_0$; the angle $\theta_0$ is considered as a small, but finite number. Using the smallness of $\theta_0$ it is easy to derive approximate expressions for all the factors that enter Eq.~\eqref{eq:16} as was done for a circular orbit in the previous section. We find ($z'<z$)
    \begin{align}\label{eq:45}
    c\tau
    &=
    \sqrt{
    (z-z')^2
    +
    (x_0(z)-x_0(z'))^2
    }
    \approx
    z-z'
    +
    \frac{\theta_0^2}{2k_w^2(z-z')}
    (\cos k_w z-\cos k_w z')^2
    ,
    \nonumber\\
    1-\vec \beta\cdot\vec\beta'
    &\approx
    \frac{1}{2}
    \theta_0^2
    \left(
    \sin k_w z
    -
    \sin k_w z'
    \right)^2
    ,\qquad
    s(z)
    \approx
    z
    \left(1+ \frac{1}{4}\theta_0^2\right)
    -
    \frac{\theta_0^2}{8k_w}
    \sin 2k_w z
    .
    \end{align}
We now substitute these expressions into Eq.~\eqref{eq:16} and replace the integration over $s$ and $s'$ by the integration over $z$ and $z'$ using $ds\approx dz$ and $ds'=dz'$. We limit the integration over $z$ by one wiggler period and divide the result by $\lambda_w$; this gives the impedance per unit length averaged over the undulator period. Finally, we replace the integration variable $z'$  by $\zeta = z-z'$. The result is\footnote{The requirement $K\gg 1$ comes from the following consideration. If one does not take the limit $v=c$ in~\eqref{eq:16}, the exponential factor $c\tau(s,s')$ should be replaced by $v\tau(s,s')$. Tracing this term to Eq.~\eqref{eq:46} gives an addition phase term $ik\zeta (1-v/c)\approx ik\zeta/2\gamma^2$ in the exponential factor. To be able to neglect this term in comparison with $ik\theta_0^2\zeta/4$ we should require $\theta_0\ll 1/\gamma$ that is $K\gg 1$.},
    \begin{align}\label{eq:46}
    Z(k)
    &=
    \frac{4iqk_w}{\pi c}
    \int_{-\lambda_w}^{\lambda_w}
    dz
    \int_0^{\infty}
    \frac{d\zeta}{\zeta}
    \sin^2 
    \left(k_w
    \frac{\zeta}{2}
    \right)
    \cos^2 
    \left(
    k_wz
    -
    \frac{k_w\zeta}{2}
    \right)
    \nonumber\\
    &\times
    \exp\left[
    -
    iq
    \left(
    k_w\zeta
    -
    \sin (k_w\zeta)
    \cos (2k_wz-k_w\zeta)
    -
    \frac{8}{k_w\zeta}
    \sin^2 
    \left(
    \frac{k_w\zeta}{2}
    \right)
    \sin^2 
    \left(
    k_wz
    -
    \frac{k_w\zeta}{2}
    \right)
    \right)
    \right]
    ,
    \end{align}
where
    \begin{align}\label{eq:47}
    q
    =
    \frac{k\theta_0^2}{4k_w}
    .
    \end{align}
Analysis of this formula (which we do not present here) shows  that this expression coincides with the result of Ref.~\cite{wu03rs}. Note that parameter $q$ is equal to the ratio of the frequency $ck$ to the fundamental radiation frequency of the wiggler $\approx 4ck_w\gamma^2/K^2$.

The limiting case $q\ll 1$ deserves a special attention---this is the case when the wavelength $2\pi/k$ is much longer the the wiggler fundamental wavelength of radiation. To calculate the real part of the impedance in this limit, Eq.~\eqref{eq:46} can be simplified taking into account that the integral converges at distances $\zeta\sim \lambdabar/\theta_0^2\gg 1$, so that we can neglect  terms on the order of 1 and $\sim 1/k_w\zeta$ in the phase and replace $\cos^2\left(k_wz-{k_w\zeta}/{2}\right)$ by its averaged value $\frac{1}{2}$,
    \begin{align}\label{eq:48}
    \Re Z(k)
    &=
    \frac{k}{c}
    \theta_0^2
    \int_0^{\infty}
    \frac{d\zeta}{\zeta}
    \sin^2 
    \left(
    \frac{k_w\zeta}{2}
    \right)
    \sin\left(
    \frac{1}{4}\theta_0^2
    k\zeta
    \right)
    =
    \frac{\pi k}{4c}
    \theta_0^2
    .
    \end{align}
Similarly, for the imaginary part we find
    \begin{align}\label{eq:49}
    \Im Z(k)
    &=
    \frac{k}{c}
    \theta_0^2
    \int_0^{\infty}
    \frac{d\zeta}{\zeta}
    \sin^2 
    \left(
    \frac{k_w\zeta}{2}
    \right)
    \cos\left(
    \frac{1}{4}\theta_0^2
    k\zeta
    \right)
    =
    -
    \frac{k}{2c}
    \theta_0^2
    \ln
    \left(\frac{1}{4k_w}\theta_0^2    k\right)
    .
    \end{align}
These results are also in agreement with~\cite{wu03rs}.

%
\section{Impedance of a kink orbit}\label{sec:VI}
%

We now proceed to the calculation of the radiation impedance for several types of orbits that have not been studied before in the literature. 

\begin{figure}[htb]
\centering
\includegraphics[width=0.6\textwidth, trim=0mm 0mm 0mm 0mm, clip]{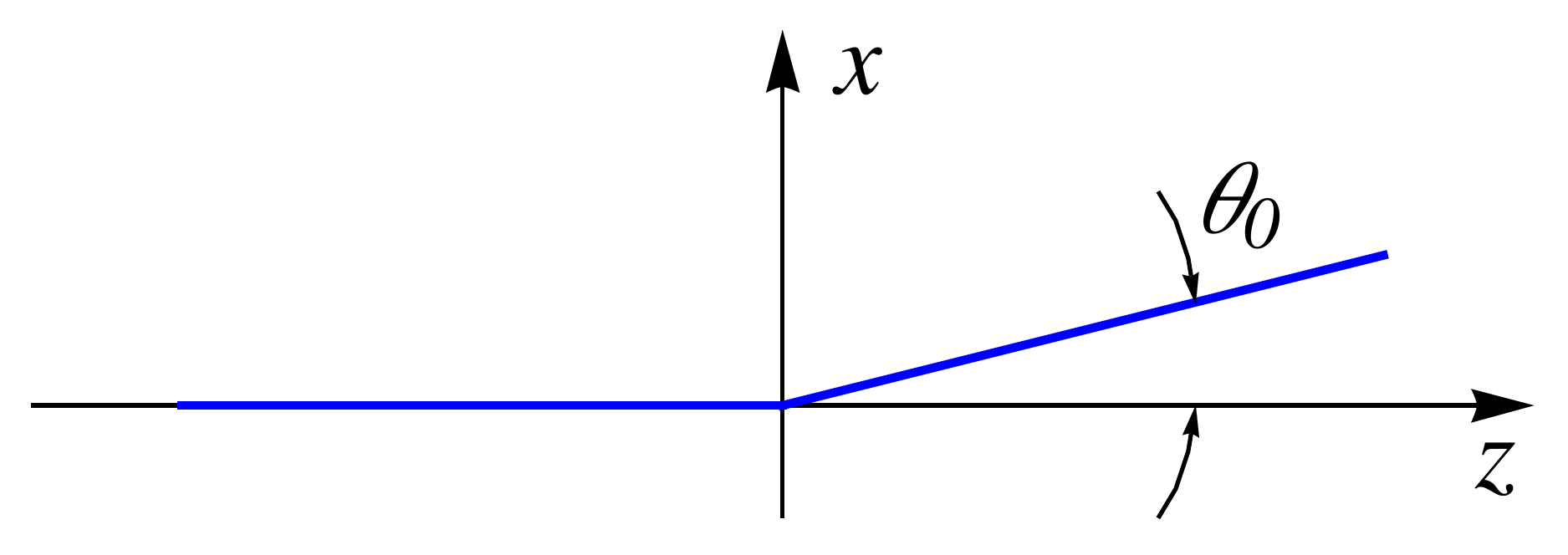}
\caption{Kink orbit shown in blue corresponds to a short magnet that deflects the orbit by a small angle $\theta_0 \ll 1$.}
\label{fig:3}
\end{figure}

One of the simplest cases is presented by a short dipole magnet that deflects the beam by angle $\theta_0\ll 1$. In our analysis we neglect the length of the magnet and consider the orbit consisting of two straight lines with the second one rotated by a small angle $\theta_0$ relative to the first: $x_0(z) = 0$ for $z<0$ and $x_0(z) = \theta_0 z$ for $z>0$, see Fig.~\ref{fig:3}.
Radiation of a point charge moving on such an orbit is studied in the textbook~\cite{jackson}---it can be related to the low-frequency limit of the bremsstrahlung radiation. A more complicated case where the finite length of the magnet is taken into account is considered in the next Section.

\begin{figure}[htb]
\centering
\includegraphics[width=0.6\textwidth, trim=0mm 0mm 0mm 0mm, clip]{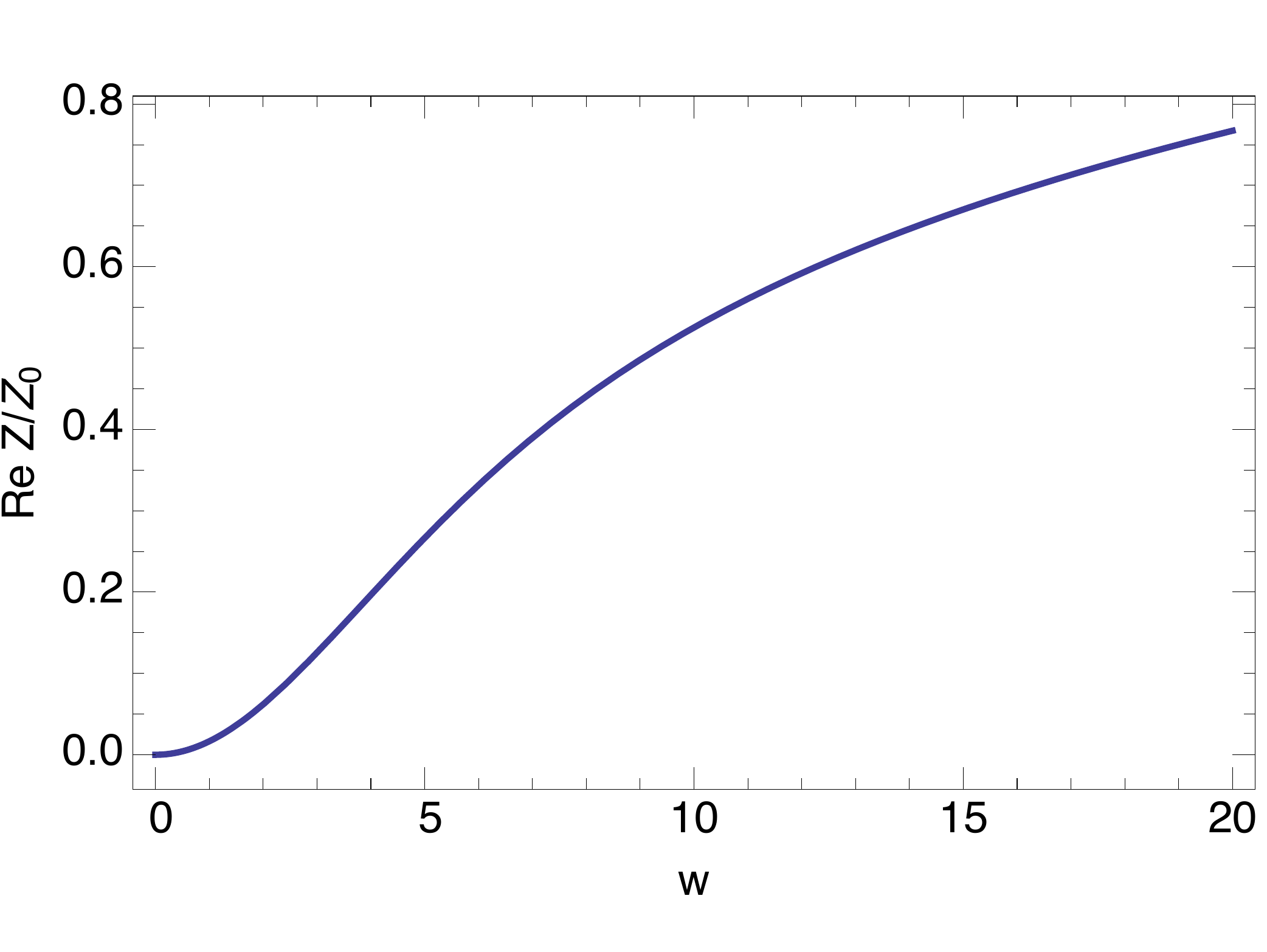}
\caption{Impedance of a kink as a function of parameter $w=ka\theta_0$.}
\label{fig:4}
\end{figure}

We will use Eq.~\eqref{eq:29} assuming a gap $a$ between the parallel conducting plates. It is clear that the integrand is not equal to zero only if $z>0$ and at the same time $z'<0$, because otherwise $1-\vec \beta(z)\cdot\vec\beta(z') = 0$. Using the smallness of  angle $\theta_0$, for $z>0$ and $z'<0$, we find
    \begin{align}\label{eq:50}
    c\tau(z,z')
    &\approx
    z-z'
    +
    \frac{1}{2(z-z')}
    \theta_0^2
    z^2
    ,\qquad
    s-s'
    =
    \sqrt{z^2+\theta_0^2 z^2}
    -
    z'
    \approx
    z-z'
    +
    \frac{1}{2}
    \theta_0^2 z
    ,
    \nonumber\\
    1-\vec \beta\cdot\vec\beta'
    &=
    \frac{1}{2}
    \theta_0^2
    .
    \end{align}
This gives the following expression for the impedance
    \begin{align}\label{eq:51}
    Z(k)
    &\approx
    (i-1)
    \frac{\sqrt{\pi k} }{ac}
    \theta_0^2
    \sum_{p=0}^\infty
    \int_{0}^\infty
    dz
    \int_{z}^\infty
    \frac{d\zeta}{\sqrt{\zeta}}
    \exp
    \left(
    -ik\frac{1}{2}\theta_0^2z    
    +
    i
    k
    \frac{1}{2\zeta}
    \theta_0^2z^2
    -
    i
    \zeta
    \frac{(2p+1)^2\pi^2}{2ka^2}
    \right)
    ,
    \end{align}
with $\zeta = z-z'$. Changing the order of integration in~\eqref{eq:51}, and using the relation
    \begin{align}\label{eq:52}
    \int_{0}^\zeta
    &dz
    \exp
    \left(
    -ik\frac{1}{2}\theta_0^2z    
    +
    i
    k
    \frac{1}{2\zeta}
    \theta_0^2z^2
    \right)
    =
    -
    \sqrt{\frac{\pi}{2}} 
    (i+1) 
    e^{-{i t^2}/{4}}
    \mathrm{erf}\left(
    \frac{i-1}{2\sqrt{2}}t\right)
    \sqrt{\frac{2\zeta}{k\theta_0^2 }}
    ,
    \end{align}
where $t=\sqrt{\zeta k\theta_0^2/2}$ and $\mathrm{erf}(x)$ is the error function, we arrive at the following equation
    \begin{align}\label{eq:53}
    Z(k)
    &=
    \frac{8\pi}{ack\theta_0}
    \sum_{p=0}^\infty
    \int_{0}^\infty
    tdt
    \exp
    \left(
    -
    i
    t^2
    \frac{(2p+1)^2\pi^2}{k^2a^2\theta_0^2}
    \right)
    e^{-{i t^2}/{4}}
	\text{erf}\left(
    \frac{i-1}{2\sqrt{2}}
    t\right)
    .
    \end{align}
Using $w=ka\theta_0$ we can write this equation in the following form
    \begin{align}\label{eq:54}
    \frac{Z(w)}{Z_0}
    &=
    \frac{2}{w}
    \sum_{p=0}^\infty
    \int_{0}^\infty
    tdt
    \exp
    \left(
    -
    i
    t^2
    \left[
    \frac{(2p+1)^2\pi^2}{w^2}
    +
    \frac{1}{4}
    \right]
    \right)
    \mathrm{erf}\left(
    \frac{i-1}{2\sqrt{2}}
    t\right)
    ,
    \end{align}
where $Z_0=4\pi/c$ is the impedance of free space. The integral on the right-hand side can be easily calculated numerically as a function of parameter $w$; the sum can also be calculated analytically 
    \begin{align}\label{eq:55}
    \frac{Z(w)}{Z_0}
    &=
    \frac{1}{2\pi }
    \left[
    {\psi ^{(0)}
    \left(
    \frac{1}{2}
    +
    \frac{iw}{4 \pi}
    \right)
    +
    \psi ^{(0)}
    \left(
    \frac{1}{2}
    -
    \frac{iw}{4 \pi}
    \right)
    -2 \psi ^{(0)}\left(\frac{1}{2}\right)}
    \right]
    ,
    \end{align}
where $\psi ^{(0)}(x)=\Gamma'(x)/\Gamma(x)$ is the polygamma function of order zero and $\Gamma(x)$ is the gamma function. The plot of this function is shown in Fig.~\ref{fig:4}. As it turns out, the impedance~\eqref{eq:54} is purely real, $\Im Z=0$.

Let us consider the limiting cases of large and small values of $w$. For $w\ll 1$, assuming $p\sim 1$, the main contribution to the integral comes from the region $t\sim w \ll 1$. In this region the error function can be replaced by its asymptotic values for $t\ll 1$, $\mathrm{erf}(x)\approx 2x/\sqrt{\pi}$. We then obtain,
    \begin{align}\label{eq:56}
    \frac{Z(w)}{Z_0}
    &\approx
    \frac{2(i-1)}{w\sqrt{2\pi}}
    \sum_{p=0}^\infty
    \int_{0}^\infty
    t^2dt
    \exp
    \left(
    -
    i
    t^2
    \frac{(2p+1)^2\pi^2}{w^2}
    \right)
    =
    \frac{w^2}{2\pi^3}
    \sum_{p=0}^\infty
    \frac{1}{(2p+1)^3}
    =
    \frac{7w^2}{16\pi^3}
    \zeta(3)
    ,
    \end{align}
where $\zeta(x)$ is the Riemann zeta function. In the opposite limit, $w\gg 1$, from the asymptotic approximation of the polygamma function, it follows that ${Z(w)}/{Z_0}\sim \ln w$.

Note that in the limit $a\to\infty$ the impedance of a kink diverges because this limit corresponds to $w\to\infty$. Hence, the radiation impedance of a kink is not defined in free space (which formally corresponds to $a=\infty$). This is of course a consequence of our assumption $v=c$.

The physical mechanism behind the radiation impedance of a kink can be attributed to the edge radiation of the beam, as discussed in Section~\ref{sec:III}. Given that the minimal transverse wavenumber $k_\perp$ in $y$ direction is equal to $\pi/a$, we conclude that the bulk of the edge radiation energy is localized at angles $\theta\sim k_\perp/k\sim\pi/ak$. There are two cones of radiation: the first one is localized around the initial direction of motion, the $z$ axis, and the second one is around the deflected direction of motion at angle $\theta_0$. The regime $w\ll 1$ corresponds to the overlapping of the edge radiation cones from the incoming and outgoing directions. The opposite regime, $w \gg 1$, corresponds to the case when the cones are well separated in space.

Understanding the physical mechanism behind the impedance allows us to estimate the formation length $l_f$ of the radiation---the distance after which radiation decouples from the charge. As usually $l_f$ is estimated at $l_f \sim 1/k\theta^2$ where $\theta$ is the angular spread of the radiation; this gives $l_f  \sim a^2 k$. Requiring the formation length to be larger than the reduces wavelength $1/k$ we obtain the condition when our analysis is correct, $ak\gg1$, which we have already formulated in Eq.~\eqref{eq:25}. In the opposite limit, one cannot truncate the integration over $s'$ by replacing the upper infinite limit by finite $s$, as was done in transition from~\eqref{eq:9} to~\eqref{eq:10}.

\begin{figure}[htb]
\centering
\includegraphics[width=0.6\textwidth, trim=0mm 0mm 0mm 0mm, clip]{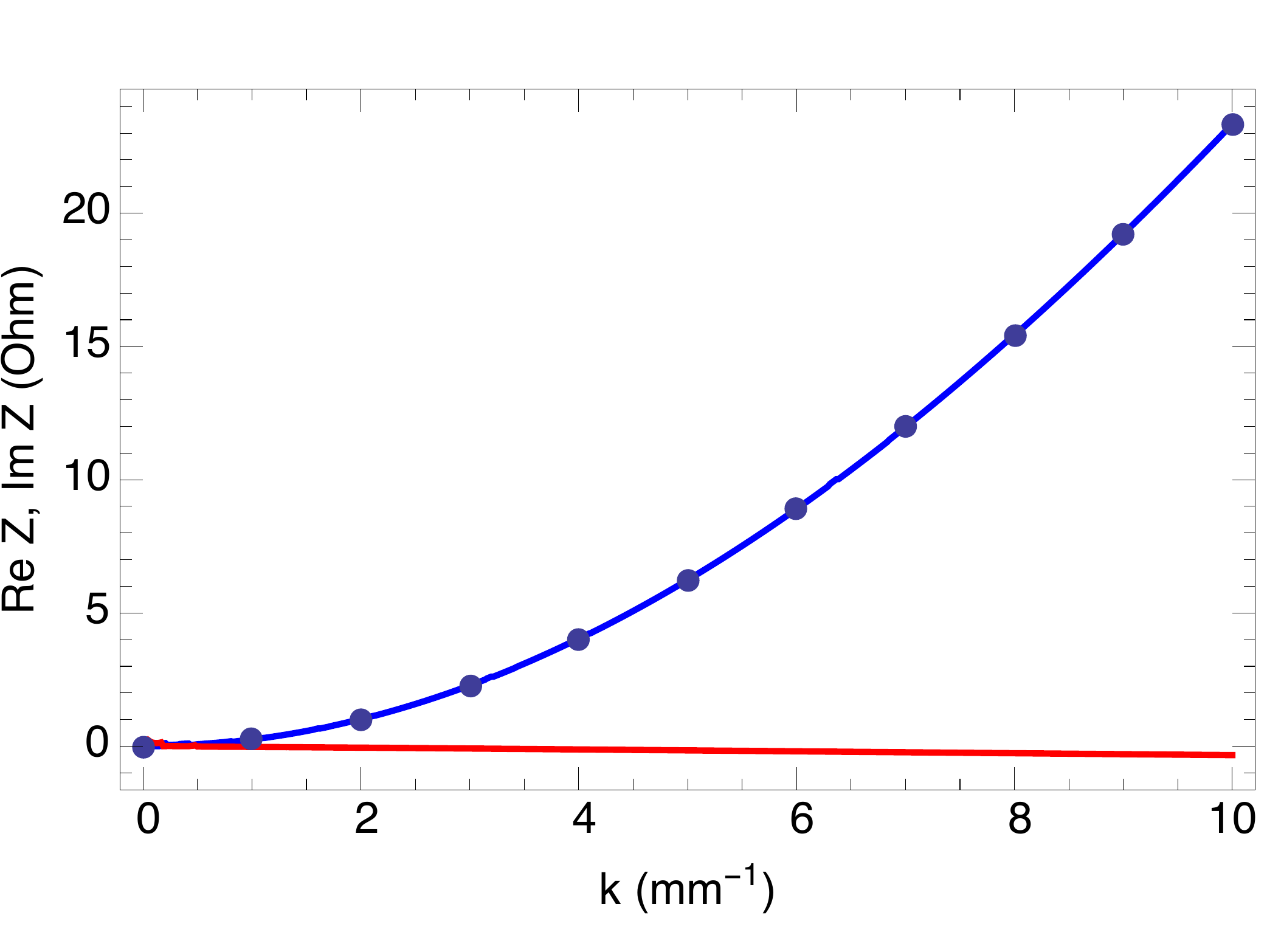}
\caption{Comparison of the analytical theory (dots) and numerical simulations (solid lines) for the case of a kink orbit. The blue line shows the real part and the red line shows the imaginary part of $Z$ computed by the code. The dots show $\Re Z$ computed using Eq.~\eqref{eq:54} (the imaginary part is equal to zero and is not shown).}
\label{fig:5}
\end{figure}

In Fig.~\ref{fig:5} 
we compare the analytical result obtained with Eq.~\eqref{eq:55} with the numerical simulation carried out  with the CSRZ code. With the code we simulated a short bending magnet of length $L=1$ cm and the bending radius of $\rho=1$ m. The vertical size of the vacuum chamber is $a=2$ cm and the aspect ratio $b/a=5$. In analytical calculations we used the same $a=2$ cm and the bending angle $\theta_0=L/\rho=0.01$. The last point on the plot corresponds to the dimensionless parameter $w=ka\theta_0=2$. Note that the numerical simulation shows a small imaginary part of $Z$; in this regard it slightly deviates from the analytical model that predicts $\Im Z=0$. The simulated real part of the impedance agrees very well with the analytical one.

%
\section{Bending magnet of finite length with shielding}\label{sec:VII}
%

\begin{figure}[htb]
    \centering
    \includegraphics[width=0.45\textwidth, trim=0mm 0mm 0mm 0mm, clip]{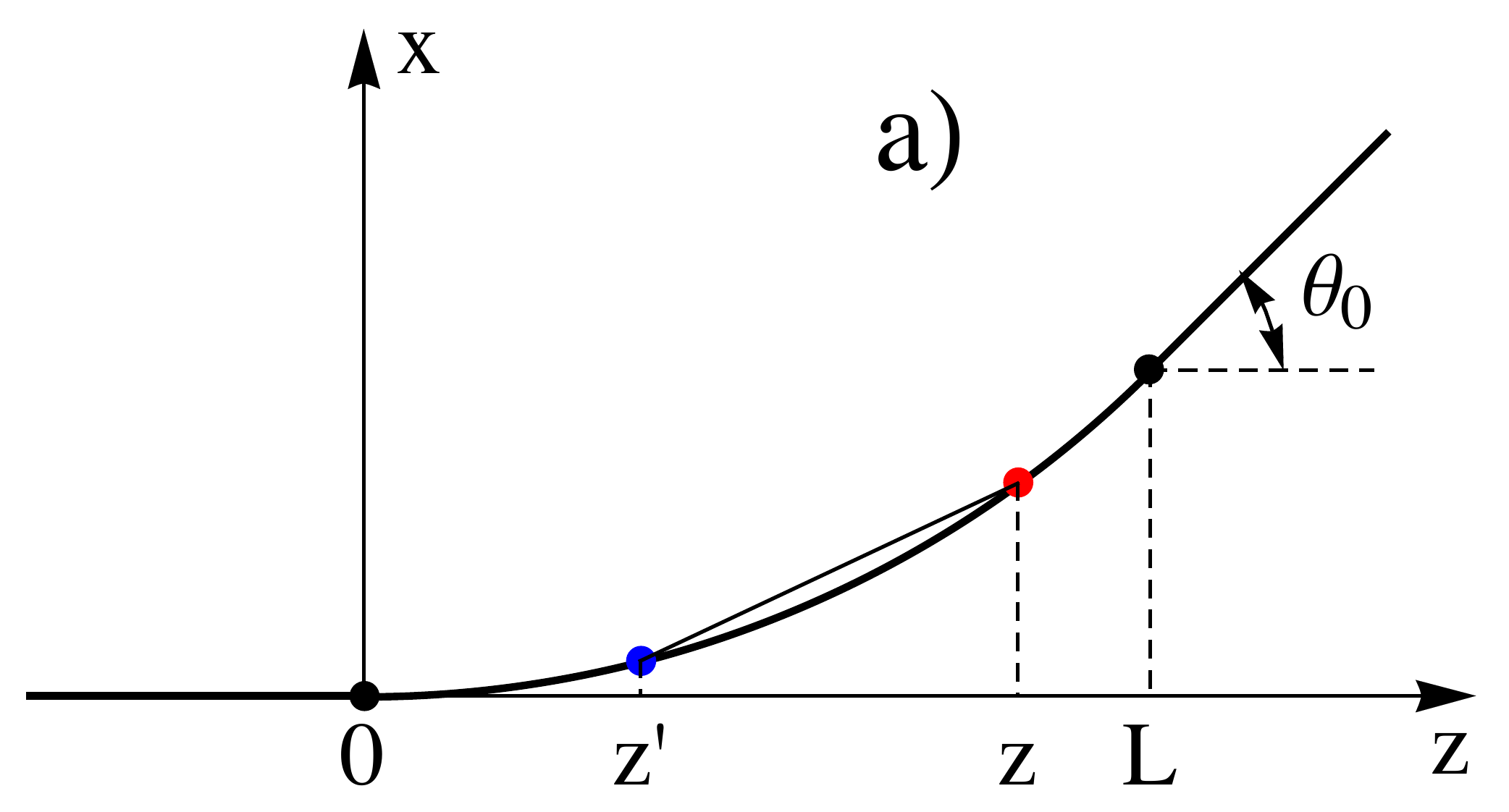}
    \includegraphics[width=0.45\textwidth, trim=0mm 0mm 0mm 0mm, clip]{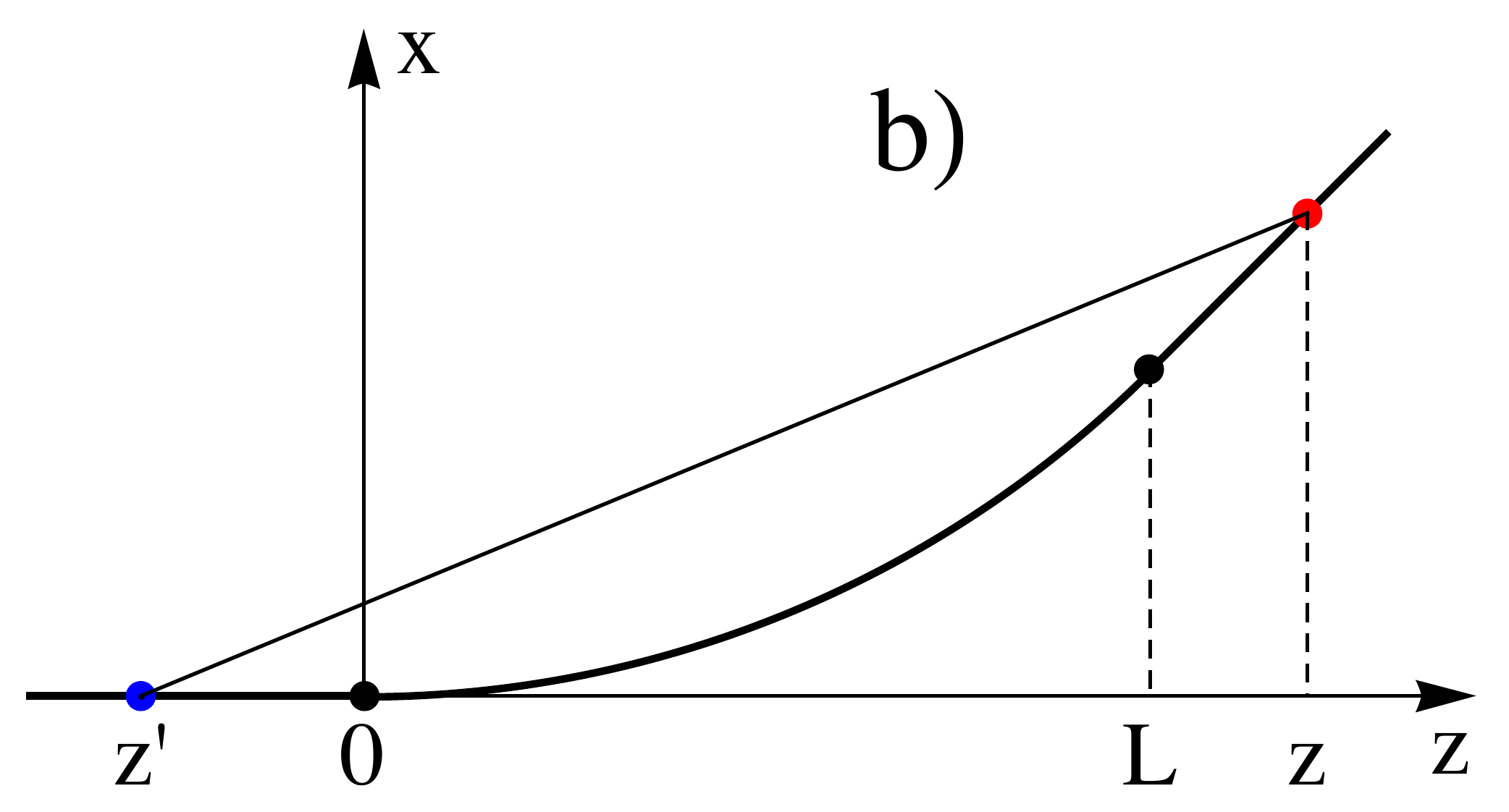}
    \includegraphics[width=0.45\textwidth, trim=0mm 0mm 0mm 0mm, clip]{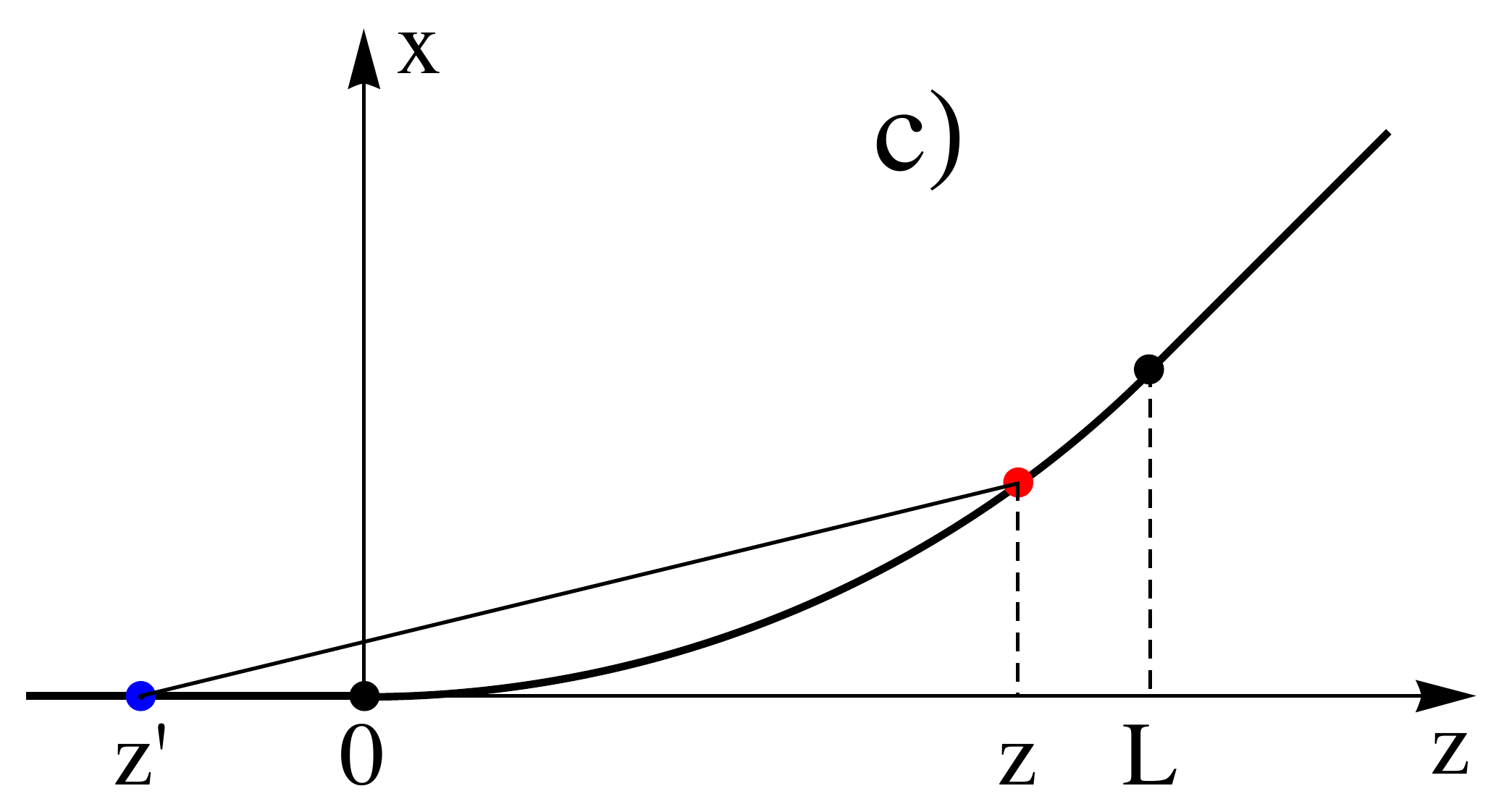}
    \includegraphics[width=0.45\textwidth, trim=0mm 0mm 0mm 0mm, clip]{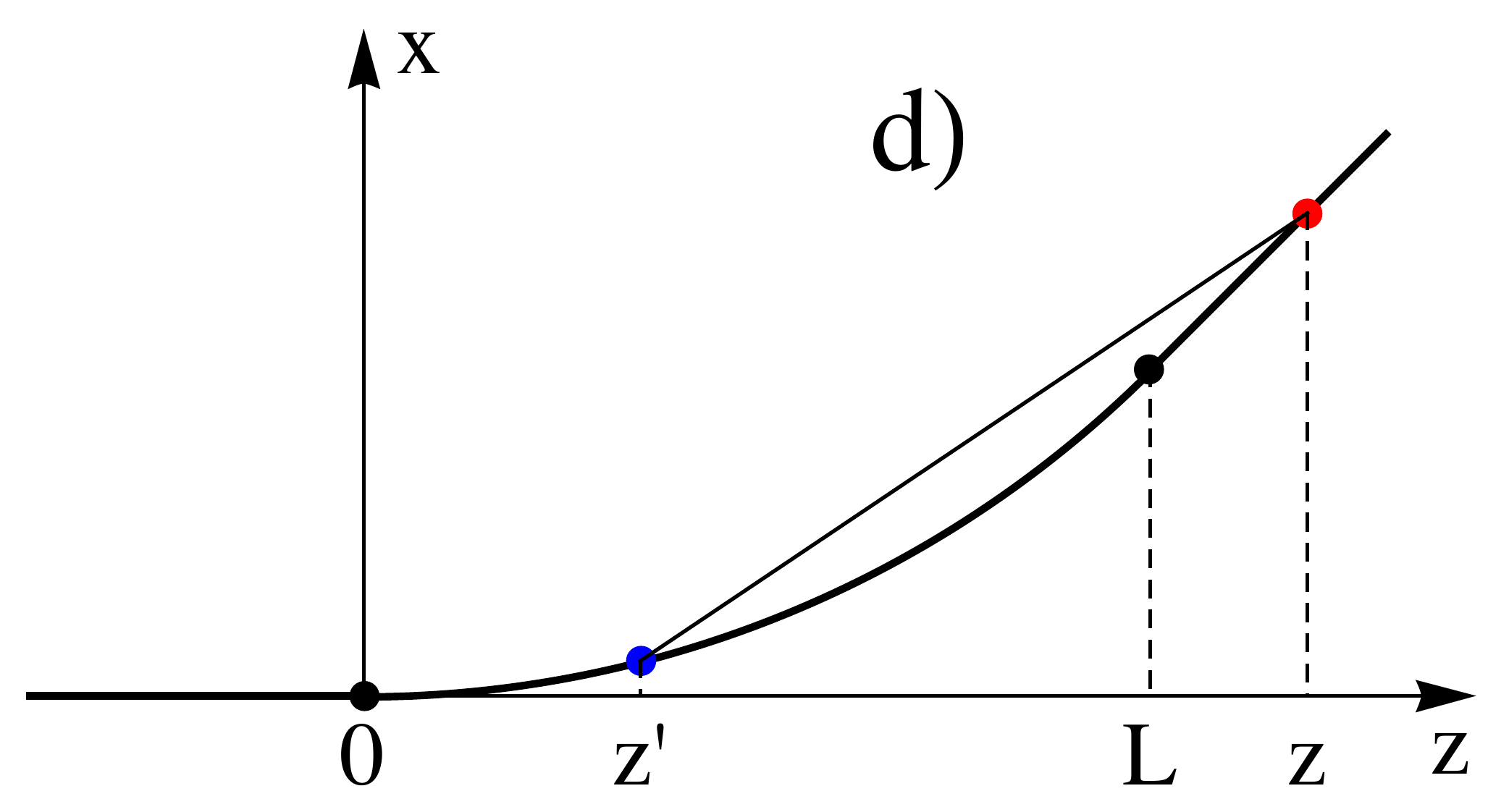}
    \caption{The orbit for a bending magnet of length $L$ consists of straight line $z<0$, a circular arc occupying the region $0<z<L$ and a straight line in the region $z>L$ tilted at angle $\theta_0$. Panes a), b), c) and d) show four different situations for relative locations of the leading point $z$ (shown by the red dot) and that of the trailing point $z'$ (shown by the blue dot). The beam moves from left to right.}
    \label{fig:6}
    \end{figure}

We  now consider a bending magnet of length $L$ and bending radius $\rho$. The magnet occupies the region $0<z<L$. The orbit is located in the midplane of two shielding parallel plates with the gap $a$ and consists of a straight line that enters the magnet at $z=0$, a circular arc inside the magnet, and a straight line exiting the magnet; see Fig.~\ref{fig:6}.  
We assume that the bending angle $\theta_0\approx L/\rho$ is small, $\theta_0\ll 1$.

Using Eq.~\eqref{eq:29} for the calculation of the impedance and remembering that $z'<z$ we will have four situations where the analytical expressions for the integrand in~\eqref{eq:29} have different forms. They are: both $z$ and $z'$ are located inside the magnet as shown in Fig.~\ref{fig:6} a; both $z$ and $z'$ are located outside of the magnet as shown in Fig.~\ref{fig:6} b; $z$ is inside and $z'$ is outside, Fig.~\ref{fig:6} c; $z$ is outside and $z'$ is inside, Fig.~\ref{fig:6} d. We denote the corresponding contributions to the impedance by $Z_1$, $Z_2$, $Z_3$ and $Z_4$, respectively; they are derived in Appendix~\ref{app:B} and given by Eqs.~\eqref{eq:68}, \eqref{eq:73}, \eqref{eq:78} and~\eqref{eq:82}. For each region we find approximate expressions for $\tau(z,z')$, $s-s'$ and $1-\vec \beta\cdot\vec\beta'$ in terms of $z$ and $z'$. It turns out that one of the integrations in~\eqref{eq:29} can be carried out analytically and the result is expressed through either elementary or special functions. The resulting expression for the impedance consists of a sum over $p$ of one dimensional integrals that can be computed numerically.

In Appendix~\ref{app:B}, we also show through a direct calculation that in free space ($a=\infty$) the contribution $Z_4$ diverges at the upper limit and the radiation impedance is infinite. This proves the statement made in Section~\ref{sec:III}.

\begin{figure}[htb]
    \centering
    \includegraphics[width=0.49\textwidth, trim=0mm 0mm 0mm 0mm, clip]{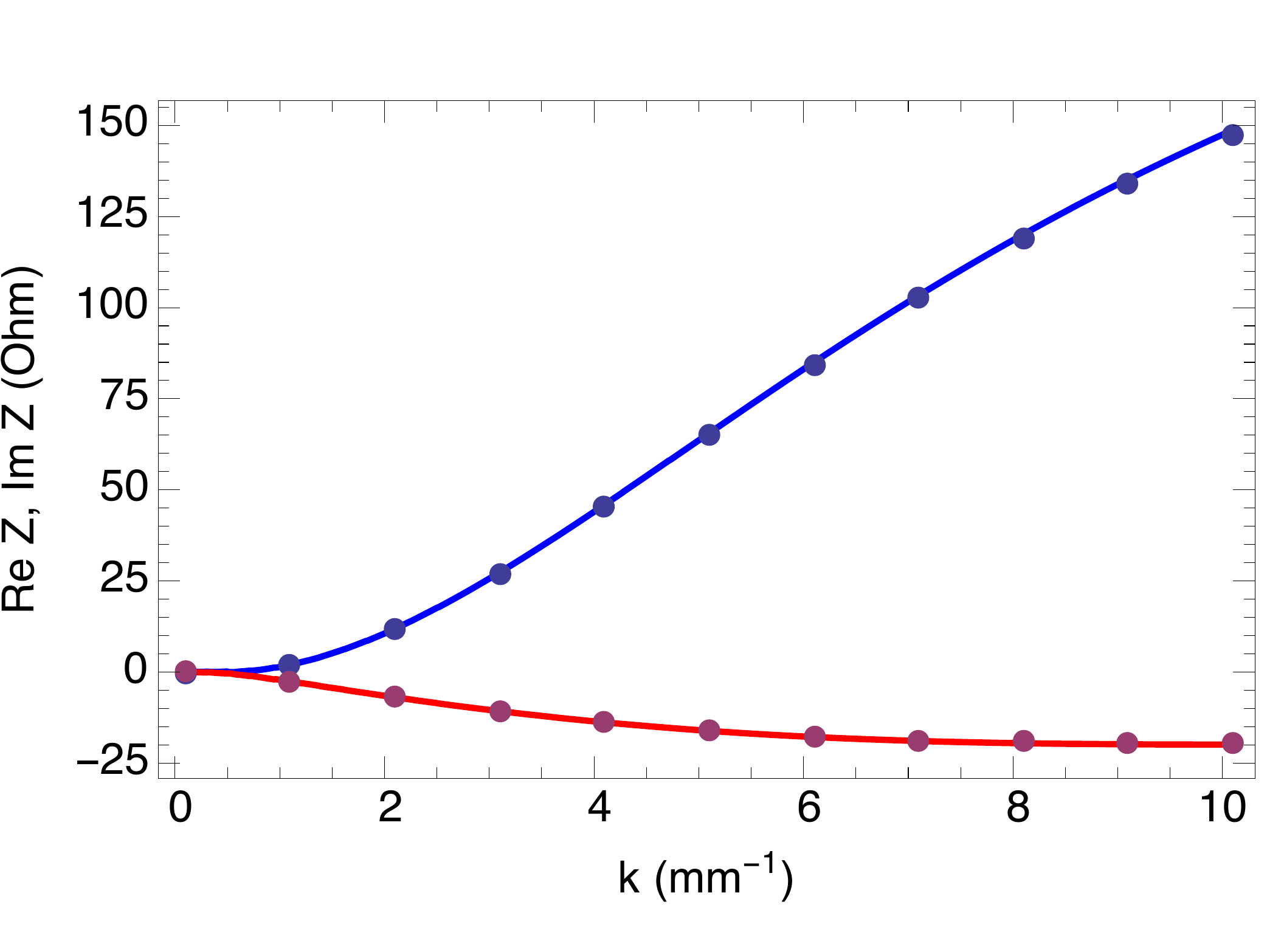}
    \includegraphics[width=0.49\textwidth, trim=0mm 0mm 0mm 0mm, clip]{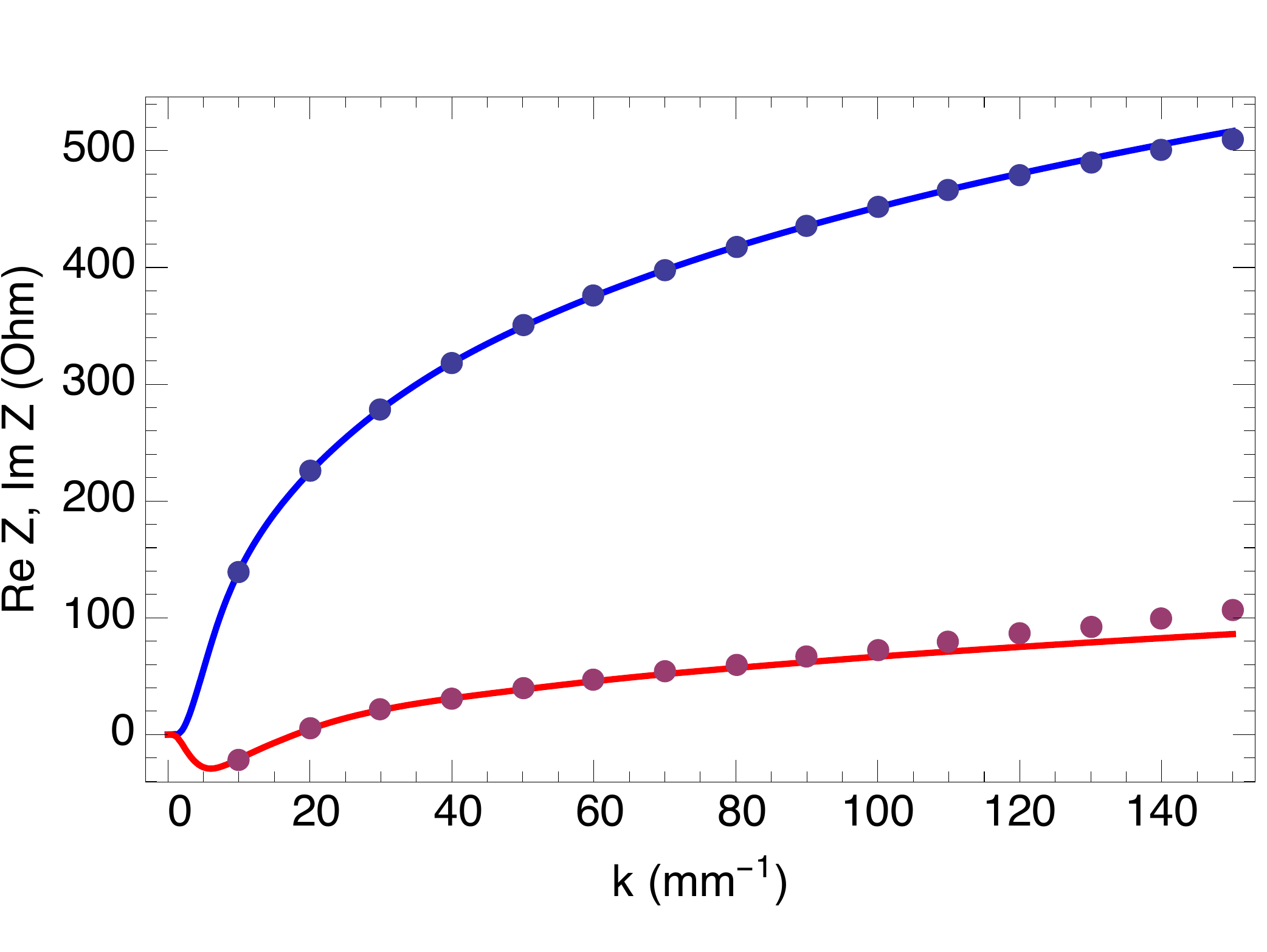}
    \caption{Comparison of analytical calculations (shown by dots) with computer simulations (shown by solid lines):  $\Re Z$ (blue) and $\Im Z$ (red).}
    \label{fig:7}
    \end{figure}

To demonstrate the capabilities of the analytical method, in Fig.~\ref{fig:7}, we benchmark our formulas with numerical simulations. 
The left pane shows the impedance for the bending magnet with $L=20$ cm, $\rho=5$ m, $a=2$ cm, and the right pane shows the impedance for a magnet with $L=55$ cm, $\rho=12.94$ m, $a=2$ cm. The second magnet has parameters of the magnets in the second bunch compressor of the LSLS-II free electron laser project~\cite{lcls-ii-csr}. In numerical calculations we used the aspect ratio 5 for the first case and 4 for the second one. The plots show an excellent agreement between the analytical and numerical results. A slight discrepancy at very large values of $k$ at the second pane is likely due to inaccuracy associated with the parabolic equation approximation.

%
\section{Radiation impedance of a wiggler of finite length}\label{sec:VIII}
%

\begin{figure}[htb]
\centering
\includegraphics[width=0.6\textwidth, trim=0mm 0mm 0mm 0mm, clip]{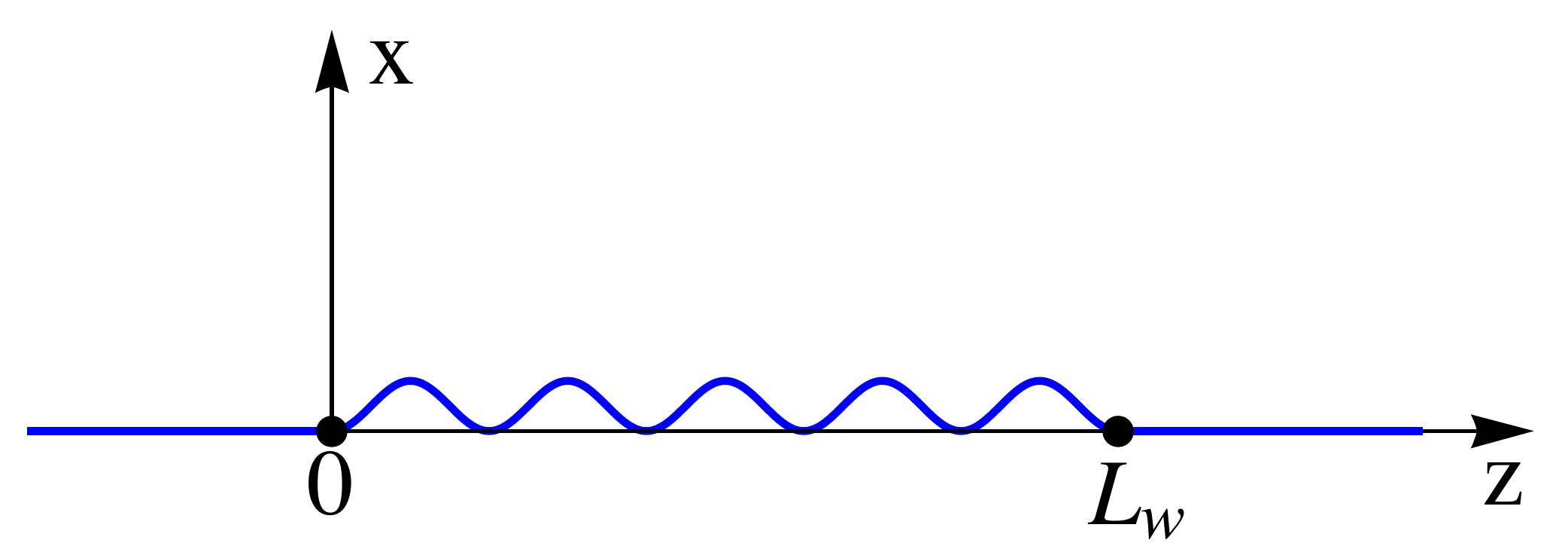}
\caption{Wiggler of length $L_w=N_w\lambda_w$ with the orbit shown by blue line.}
\label{fig:8}
\end{figure}

We now consider a plane wiggler that has $N_w$ periods ($N_w$ is an integer) with the period length $\lambda_w$ and the undulator parameter $K\gg 1$. As in Section~\ref{sec:VB}, we introduce  $\theta_0 = K/\gamma \ll 1$ and $k_w=2\pi/\lambda_w$. Particle orbits inside the wiggler, $0<z<N_w\lambda_w$, are given by the following equations
    \begin{align}
    \label{eq:57}
    x_0(z)
    &=
    \theta_0k_w^{-1}(1-\cos(k_w z))
    ,
    \qquad
    y_0(z)
    =
    0
    \nonumber\\
    \vec{\beta}_\perp(z)
    &=
    \theta_0
    \hat{\vec{x}}\sin (k_w z)
    ,\qquad
    \beta_z
    =
    1
    -
    \frac{1}{2}
    \theta_0^2
    \sin^2 (2k_w z)
    ;
    \end{align}	                                                                                                                                                                                                                                                                                                                                                                                                                                                                                                                                                                                                                                                                                                                                                                                                                                                                                                                                 
outside of the wiggler we have $x_0(z)=y_0(z)=0$. In comparison with Eqs.~\eqref{eq:43} and~\eqref{eq:44} we added inside the wiggler a constant shift $\theta_0k_w^{-1}$ to $x$ to eliminate a jump in the first derivative of the orbit in the transition  from the straight sections. The orbit is sketched in Fig.~\ref{fig:8}.

Calculating the impedance with Eq.~\eqref{eq:29} we split the contributions to $Z$ into three parts: first, $Z_1$, when $0<z,z'<L_w$; second, $Z_2$, when $-\infty<z'<0$ and $0<z<L_w$; and third, $Z_3$, corresponding to the integration $0<z'<L_w$ and $L_w<z<\infty$. The details of the calculations can be found in Appendix~\ref{app:C} with the resulting expression for the impedances given by Eqs.~\eqref{eq:92}, \eqref{eq:96} and~\eqref{eq:100}. 

\begin{figure}[htb]
\centering
\includegraphics[width=0.6\textwidth, trim=0mm 0mm 0mm 0mm, clip]{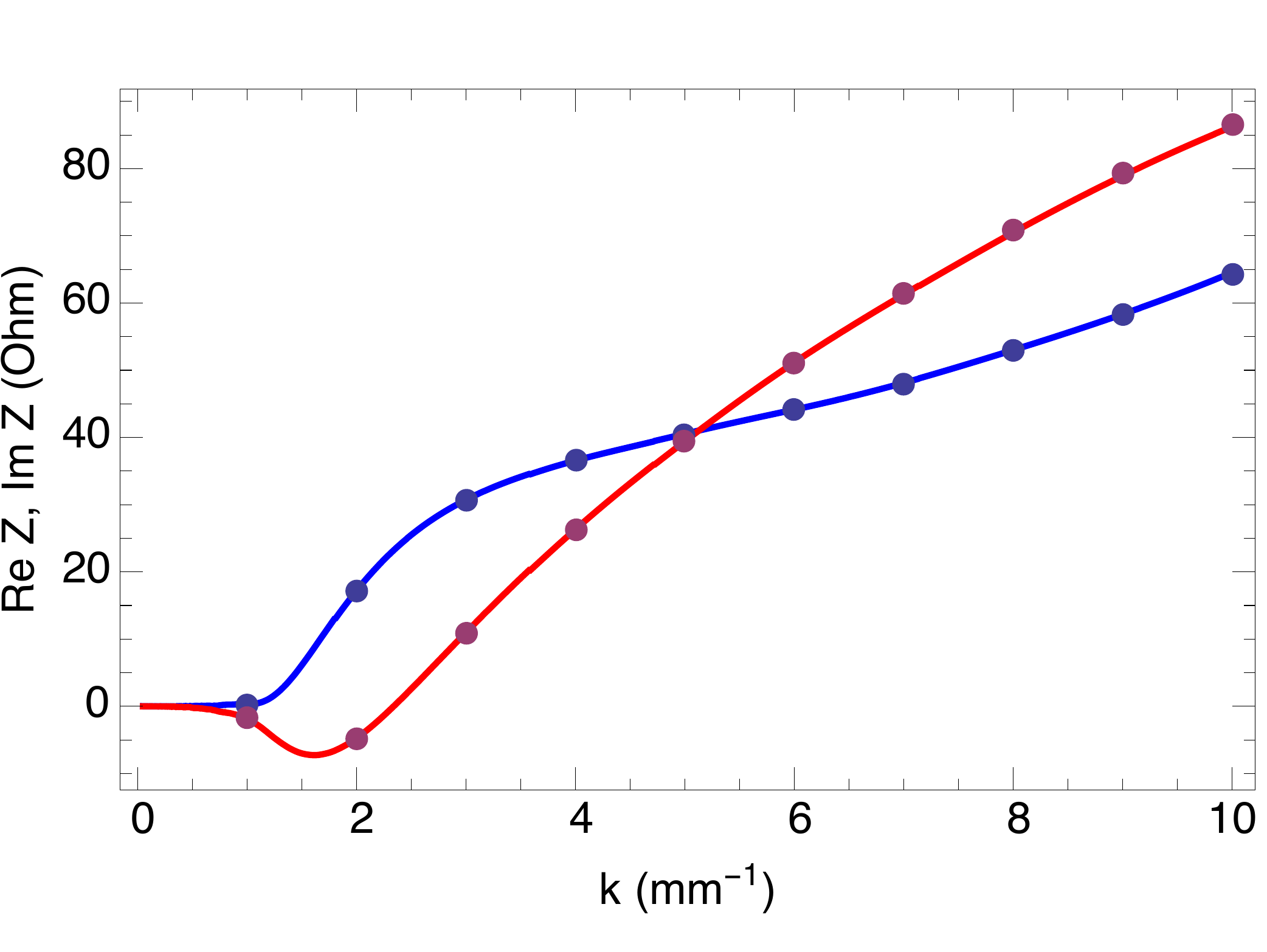}
\caption{Wiggler impedance: comparison of analytical calculations (shown by dots) with computer simulations (shown by solid lines).  $\Re Z$ is shown by blue color and $\Im Z$ is red.}
\label{fig:9}
\end{figure}

Comparison of analytical calculations with numerical simulations for a wiggler is shown in Fig.~\ref{fig:9}.
The wiggler has one period, $N_w=1$, the period length $\lambda_w=1$ m and the angle $\theta_0 = 1.6\times 10^{-2}$. The gap is $a=2$ cm and the aspect ratio $b/a=5$. In another run we also used the aspect ratio $b/a=10$---the result was the same as for $b/a=5$. We find an excellent agreement between the numerical and analytical calculations in this case too.

%
\section{Wiggler of infinite length with shielding}\label{sec:IX}
%

For a long wiggler with many periods one can use an approximation $N_w\to\infty$ and calculate the impedance averaged over one period, as it was done in Section~\ref{sec:VB} for an infinitely long wiggler in free space. This calculation is carried out in Appendix~\ref{app:D}, with the impedance given by Eq.~\eqref{eq:104}.

\begin{figure}[htb]
\centering
\includegraphics[width=0.49\textwidth, trim=0mm 0mm 0mm 0mm, clip]{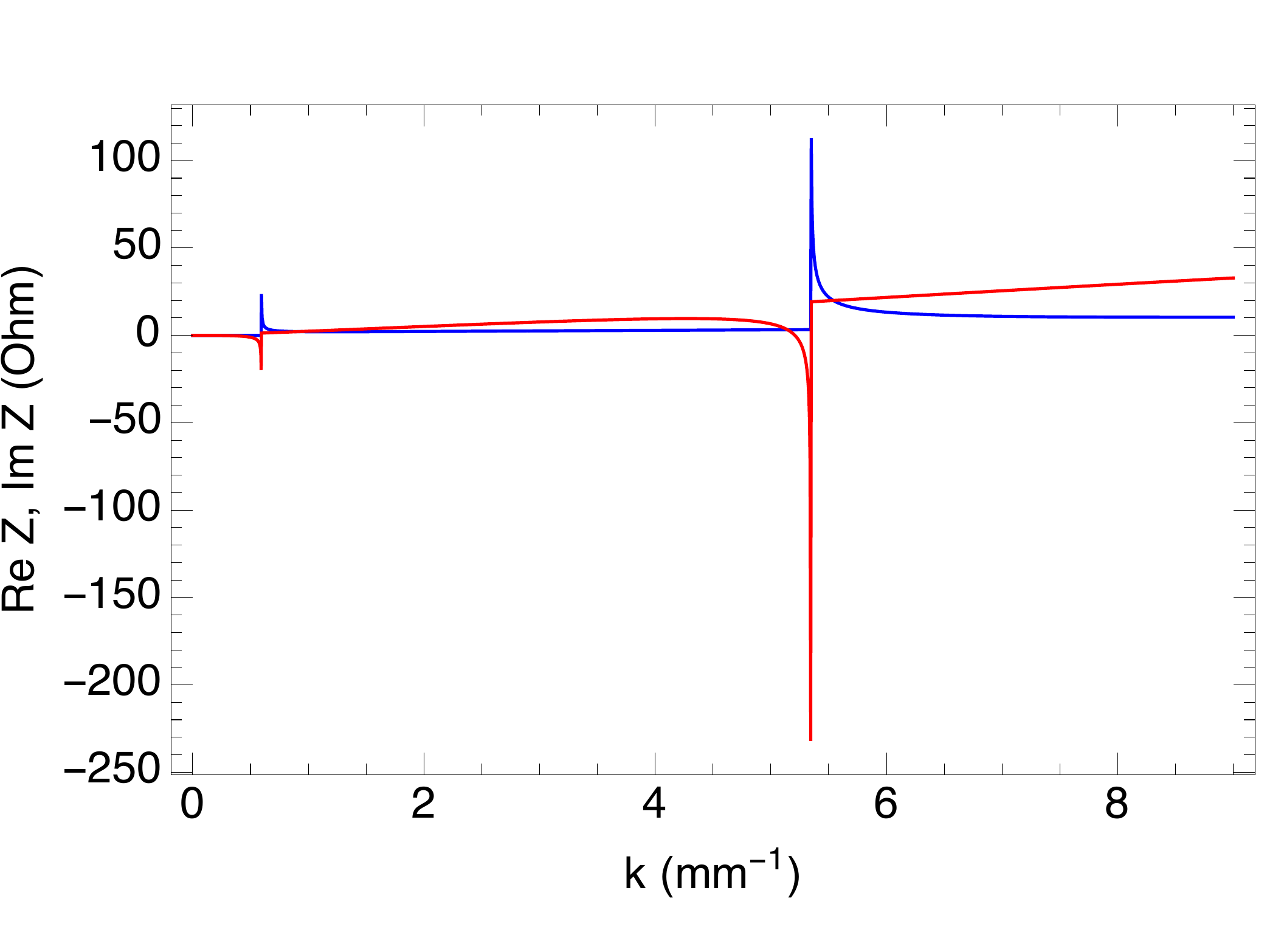}
\includegraphics[width=0.49\textwidth, trim=0mm 0mm 0mm 0mm, clip]{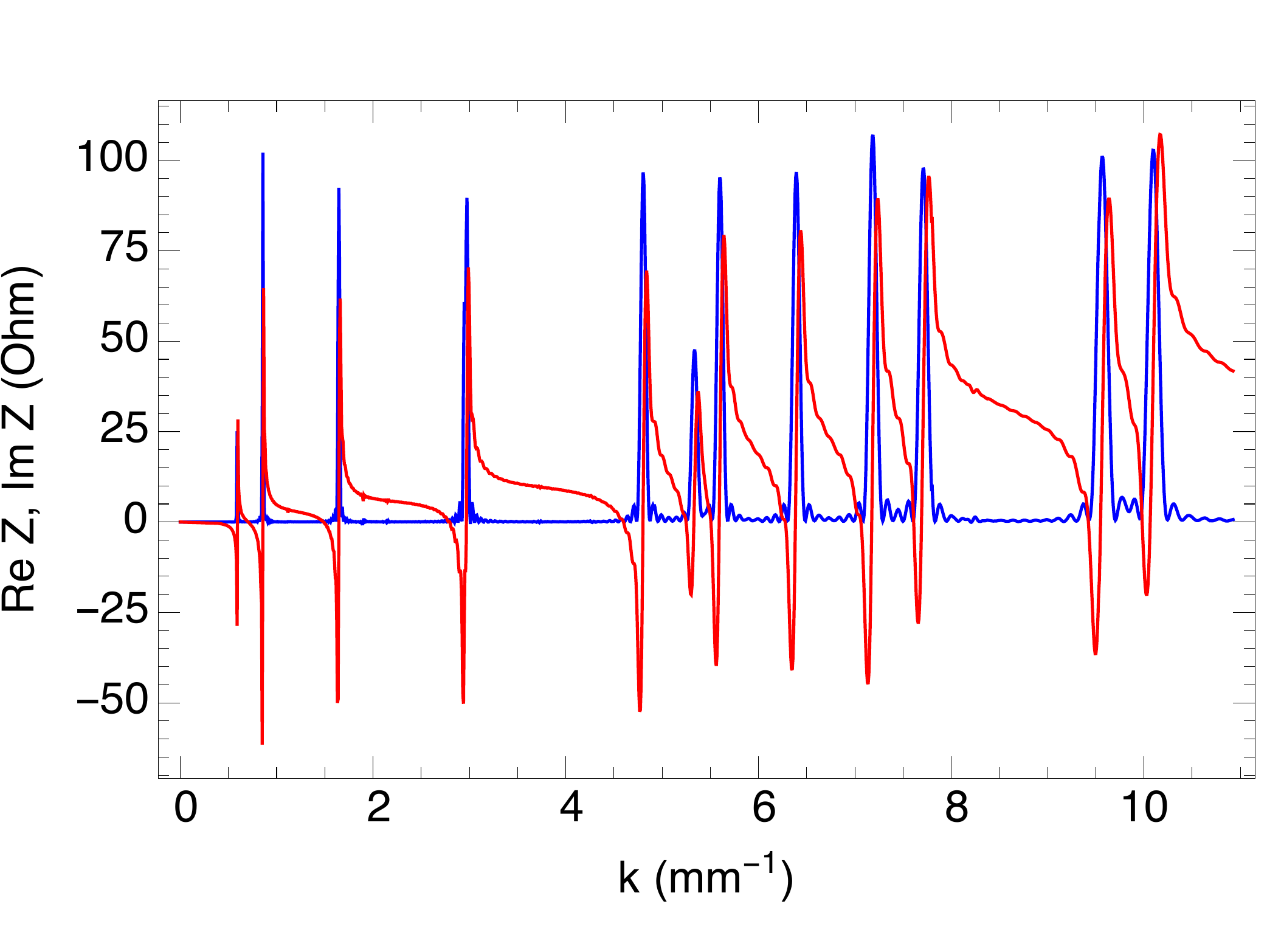}
\caption{Comparison of analytical calculations (left pane) with computer simulations (right pane) for NSLS-II wiggler impedance:  $\Re Z$ (blue) and $\Im Z$ (red).}
\label{fig:10}
\end{figure}

To test Eq.~\eqref{eq:104} we calculated the radiation impedance for NSLS-II damping wigglers~\cite{nslsII07}. The wiggler has the following parameters: $N_w=70$, $\lambda_w=10$ cm, $K=16.8$. With the  NSLS-II beam energy of 3 GeV the maximal deflection angle is $\theta_0 = 1.86\times 10^{-3}$. The vertical transverse size of the vacuum chamber $a=11.5$ mm was used for the gap between the parallel conducting plates in the analytical model. In numerical calculations the horizontal size of the vacuum chamber was taken to be three times larger than the vertical one, $b=3a$.  The impedance calculated with Eq.~\eqref{eq:104} and with the code CSRZ is shown in Fig.~\ref{fig:10}.
As one can see from this figure, the impedance is dominated by sharp, resonant-like spikes at several frequencies. The locations of the spikes is explained by the synchronicity between waveguide modes of the rectangular vacuum chamber with the wiggling trajectory of the beam~\cite{cur_preprint}. The resonant values of $k$ are defined by the following equation,
    \begin{align}\label{eq:58}
    k-k_w
    =
    \sqrt{
    k^2
    -
    \frac{\pi^2 n^2}{a^2}
    -
    \frac{\pi^2 m^2}{b^2}
    }
    ,
    \end{align} 
where $n$ is an odd and $m$ is an even number, and $a$ and $b$ are the dimensions of the rectangular cross section. From the analysis of this equations if follows that the twelve spikes on the right plot of Fig.~\ref{fig:10} are all explained by the resonances with $n=1,3$ and $m\le12$. In the parallel plates model, formally $b=\infty$, and the resonant modes are given by~\eqref{eq:58} with $m=0$. The two spikes on the left plot of Fig.~\ref{fig:10} are the $n=1,3$ resonances.
         
\begin{figure}[htb]
\centering
\includegraphics[width=0.6\textwidth, trim=0mm 0mm 0mm 0mm, clip]{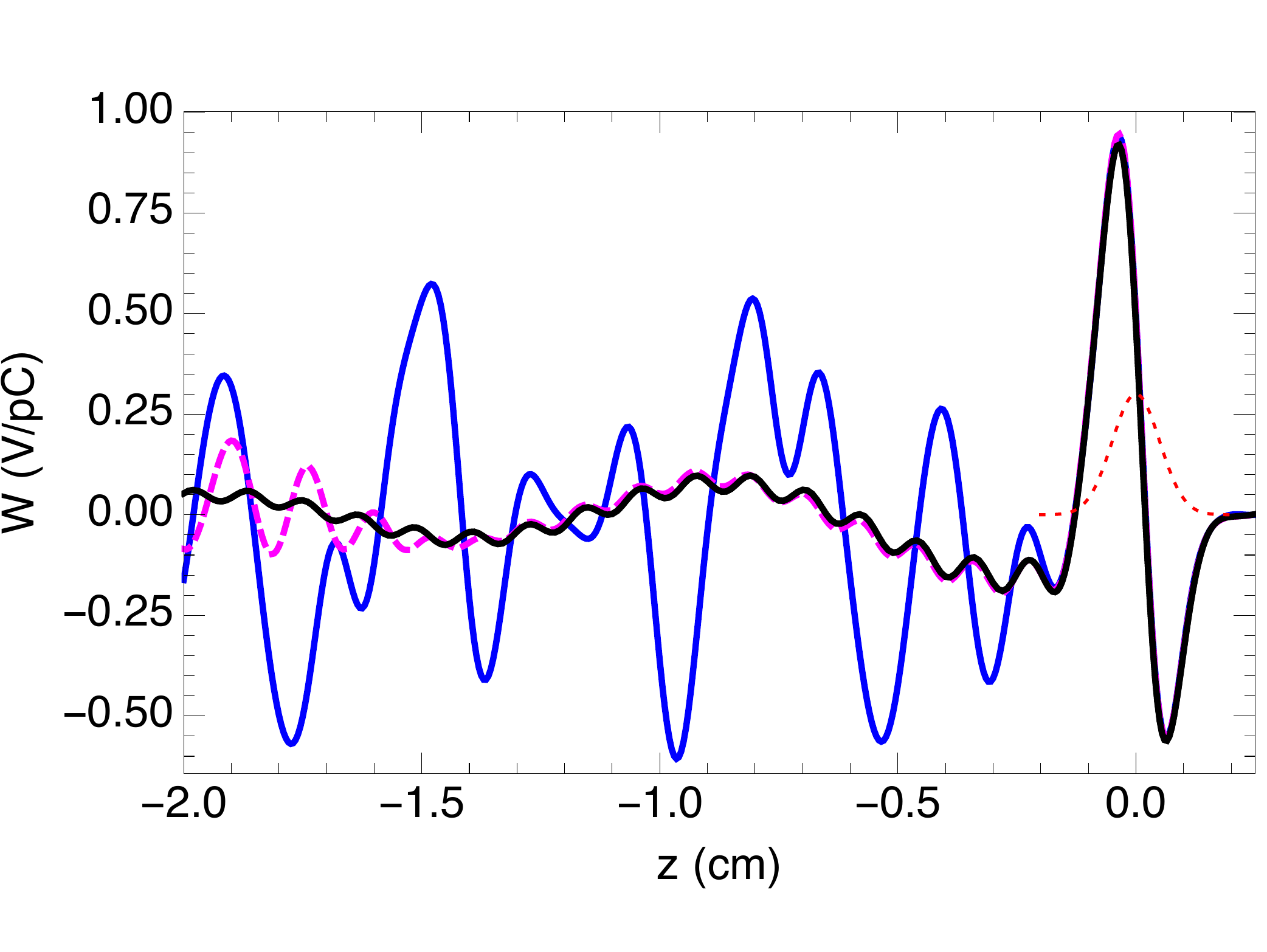}
\caption{Wakefields for a Gaussian bunch with $\sigma_z=0.5$ mm (the bunch profile is shown by red dotted line): calculated with CSRZ (blue line) and with the analytical model (red line) using impedances shown in Fig.~\ref{fig:10}. The dashed magenta shows the numerically calculated wake for the aspect ratio $b/a=16$. The bunch head is to the right; positive wake corresponds to the energy loss.
}
\label{fig:11}
\end{figure}

While impedances in the left and the right plots look very different, it is remarkable that at a short distance they correspond to the same wakefield. This is illustrated by Fig.~\ref{fig:11} in which the blue and black lines show two wakefields, numerical and analytical, calculated from the impedances shown in Fig.~\ref{fig:10} for a Gaussian bunch with rms length of 0.5 mm.
We see that the complicated resonant structure of the impedances causes deviation of the wakes at distances $z\lesssim -0.2$ cm, while for $z\gtrsim -0.2$ cm we have an excellent agreement between the wakes computed with both methods. In Fig.~\ref{fig:11} we also show another numerically calculated wakefield, for an aspect ratio $b/a=16$ (the dashed magenta line). As expected, this wake agrees much better with the analytical wake (the black line) of the parallel plates model.

%
\section{Summary}\label{sec:X}
%

In this paper, we presented general expressions, Eqs.~\eqref{eq:24} and~\eqref{eq:29}, for the radiation impedance of a relativistic beam moving on an arbitrary plane orbit between two parallel conducting plates. In the derivation of these expressions we assumed that the transverse size of the beam is infinitely small and the particles move with $v=c$. Eq.~\eqref{eq:29} additionally assumes a short bunch and an orbit that does not deviate much from the direction of the $z$ axis.

We showed that all known in the literature analytical results for the radiation impedance can be straightforwardly obtained from these expression. New analytical results were derived for  the radiation impedance with shielding for the following orbits: a kink, a bending magnet, a wiggler of finite length, and an infinitely long wiggler. All our formulas are benchmarked agains numerical simulations with the CSRZ code.

%
\section{Acknowledgements}\label{sec:XI}
%

DZ would like to thank T. Agoh, K. Ohmi and K. Yokoya for helpful discussion during the development of CSRZ code. GS would like to thank K. Ohmi for the hospitality during his visits to KEK when this work has been carried out.
 
\bibliography{\string~/gsfiles/Bibliography/books%
              ,\string~/gsfiles/bibliography/accel%
              ,\string~/gsfiles/bibliography/stupakov%
              }

\begin{thebibliography}{10}
\providecommand*{\bibinfo}[2]{#2}
\providecommand*{\eprint}[1]{#1}
\providecommand*{\url}[1]{#1}
\bibitem{derbenev95rss}
\bibinfo{author}{Y.~S. Derbenev}, \bibinfo{author}{J.~Rossbach},
  \bibinfo{author}{E.~L. Saldin}, and \bibinfo{author}{V.~D. Shiltsev},
  \bibinfo{title}{\emph{Microbunch Radiative Tail-Head Interaction}}, DESY FEL
  Report TESLA-FEL 95-05, Deutsches Elektronen-Synchrotron, Hamburg, Germany
  (\bibinfo{date}{September 1995}).
\bibitem{murphy97kg}
\bibinfo{author}{J.~B. Murphy}, \bibinfo{author}{S.~Krinsky}, and
  \bibinfo{author}{R.~L. Gluckstern}, \bibinfo{journal}{Part. Accel.}
  \bibinfo{volume}{\textbf{57}}, \bibinfo{pages}{9} (\bibinfo{date}{1997}).
\bibitem{warnock90m}
\bibinfo{author}{R.~L. Warnock} and \bibinfo{author}{P.~Morton},
  \bibinfo{journal}{Part. Accel.} \bibinfo{volume}{\textbf{25}},
  \bibinfo{pages}{113} (\bibinfo{date}{1990}).
\bibitem{ng90}
\bibinfo{author}{K.-Y. Ng}, \bibinfo{journal}{Part. Accel.}
  \bibinfo{volume}{\textbf{25}}, \bibinfo{pages}{153} (\bibinfo{date}{1990}).
\bibitem{saldin97sy}
\bibinfo{author}{E.~L. Saldin}, \bibinfo{author}{E.~A. Schneidmiller}, and
  \bibinfo{author}{M.~V. Yurkov}, \bibinfo{journal}{Nuclear Instruments and
  Methods} \bibinfo{volume}{\textbf{A398}}, \bibinfo{pages}{373}
  (\bibinfo{date}{1997}).
\bibitem{stupakov02e}
\bibinfo{author}{G.~Stupakov} and \bibinfo{author}{P.~Emma}, in
  \emph{Proceedings of 8th European Particle Accelerator Conference}, Paris,
  France (\bibinfo{date}{2002}), \bibinfo{pages}{p. 1479}.
\bibitem{borland00}
\bibinfo{author}{M.~Borland}, \bibinfo{title}{\emph{{elegant}: A Flexible
  {SDDS}-Compliant Code for Accelerator Simulation}}, Tech. Rep. LS-287,
  Argonne National Laboratory (\bibinfo{date}{2000}).
\bibitem{saldin98sy}
\bibinfo{author}{E.~L. Saldin}, \bibinfo{author}{E.~A. Schneidmiller}, and
  \bibinfo{author}{M.~V. Yurkov}, \bibinfo{journal}{Nuclear Instruments and
  Methods, Sec. A} \bibinfo{volume}{\textbf{417}}, \bibinfo{pages}{158}
  (\bibinfo{date}{1998}).
\bibitem{wu03rs}
\bibinfo{author}{J.~Wu}, \bibinfo{author}{T.~Raubenheimer}, and
  \bibinfo{author}{G.~Stupakov}, \bibinfo{journal}{Phys. Rev. ST Accel. Beams}
  \bibinfo{volume}{\textbf{6}}, \bibinfo{pages}{040701} (\bibinfo{date}{2003}).
\bibitem{bassi06etal}
\bibinfo{author}{G.~Bassi}, \bibinfo{author}{T.~Agoh},
  \bibinfo{author}{M.~Dohlus}, \bibinfo{author}{L.~Giannessi},
  \bibinfo{author}{R.~Hajima}, \bibinfo{author}{A.~Kabel},
  \bibinfo{author}{T.~Limberg}, and \bibinfo{author}{M.~Quattromini},
  \bibinfo{journal}{Nucl. Instrum. Meth.} \bibinfo{volume}{\textbf{A557}},
  \bibinfo{pages}{189} (\bibinfo{date}{2006}).
\bibitem{li08}
\bibinfo{author}{R.~Li}, \bibinfo{journal}{Phys. Rev. ST Accel. Beams}
  \bibinfo{volume}{\textbf{11}}, \bibinfo{pages}{024401}
  (\bibinfo{date}{2008}).
\bibitem{mayes09}
\bibinfo{author}{C.~Mayes} and \bibinfo{author}{G.~Hoffstaetter},
  \bibinfo{journal}{Phys. Rev. ST Accel. Beams}
  \bibinfo{volume}{\textbf{12}}(2), \bibinfo{pages}{024401}
  (\bibinfo{date}{2009}).
\bibitem{PhysRevSTAB.12.040703}
\bibinfo{author}{D.~Sagan}, \bibinfo{author}{G.~Hoffstaetter},
  \bibinfo{author}{C.~Mayes}, and \bibinfo{author}{U.~Sae-Ueng},
  \bibinfo{journal}{Phys. Rev. ST Accel. Beams} \bibinfo{volume}{\textbf{12}},
  \bibinfo{pages}{040703} (\bibinfo{date}{2009}).
\bibitem{landau_lifshitz_ctf}
\bibinfo{author}{L.~D. Landau} and \bibinfo{author}{E.~M. Lifshitz},
  \bibinfo{title}{\emph{The Classical Theory of Fields}},
  \bibinfo{volume}{vol.~2 of \emph{Course of Theoretical Physics}}
  (\bibinfo{publisher}{Pergamon}, London, \bibinfo{year}{1979}), 4th ed.,
  (Translated from the Russian).
\bibitem{bane02c}
\bibinfo{author}{K.~L.~F. Bane} and \bibinfo{author}{A.~W. Chao},
  \bibinfo{journal}{Phys. Rev. ST Accel. Beams} \bibinfo{volume}{\textbf{5}},
  \bibinfo{pages}{104401} (\bibinfo{date}{2002}).
\bibitem{chao93}
\bibinfo{author}{A.~W. Chao}, \bibinfo{title}{\emph{Physics of Collective Beam
  Instabilities in High Energy Accelerators}} (\bibinfo{publisher}{Wiley}, New
  York, \bibinfo{year}{1993}).
\bibitem{stupakov02h}
\bibinfo{author}{G.~Stupakov} and \bibinfo{author}{S.~Heifets},
  \bibinfo{journal}{Phys. Rev. ST Accel. Beams} \bibinfo{volume}{\textbf{5}},
  \bibinfo{pages}{054402} (\bibinfo{date}{2002}).
\bibitem{bosch98}
\bibinfo{author}{R.~A. Bosch}, \bibinfo{journal}{Il Nuovo Cimento}
  \bibinfo{volume}{\textbf{20 D}}, \bibinfo{pages}{483} (\bibinfo{date}{1998}).
\bibitem{Zhou2012}
\bibinfo{author}{D.~Zhou}, \bibinfo{author}{K.~Ohmi},
  \bibinfo{author}{K.~Oide}, \bibinfo{author}{L.~Zang}, and
  \bibinfo{author}{G.~Stupakov}, \bibinfo{journal}{Japanese Journal of Applied
  Physics} \bibinfo{volume}{\textbf{51}}(1R), \bibinfo{pages}{016401}
  (\bibinfo{date}{2012}).
\bibitem{stupakov03k}
\bibinfo{author}{G.~V. Stupakov} and \bibinfo{author}{I.~A. Kotelnikov},
  \bibinfo{journal}{Phys. Rev. ST Accel. Beams} \bibinfo{volume}{\textbf{6}},
  \bibinfo{pages}{034401} (\bibinfo{date}{2003}).
\bibitem{agoh04}
\bibinfo{author}{T.~Agoh}, \bibinfo{title}{\emph{Dynamics of Coherent
  Synchrotron Radiation by Paraxial Approximation}}, Ph.D. thesis, Department
  of Physics, University of Tokyo (\bibinfo{date}{2004}).
\bibitem{agoh04y}
\bibinfo{author}{T.~Agoh} and \bibinfo{author}{K.~Yokoya},
  \bibinfo{journal}{Phys. Rev. ST Accel. Beams} \bibinfo{volume}{\textbf{7}},
  \bibinfo{pages}{054403} (\bibinfo{date}{2004}).
\bibitem{wu03rsh}
\bibinfo{author}{J.~Wu}, \bibinfo{author}{T.~O. Raubenheimer},
  \bibinfo{author}{G.~V. Stupakov}, and \bibinfo{author}{Z.~Huang},
  \bibinfo{journal}{Phys. Rev. ST Accel. Beams} \bibinfo{volume}{\textbf{6}},
  \bibinfo{pages}{104404} (\bibinfo{date}{2003}).
\bibitem{jackson}
\bibinfo{author}{J.~D. Jackson}, \bibinfo{title}{\emph{Classical
  Electrodynamics}} (\bibinfo{publisher}{Wiley}, New~York,
  \bibinfo{year}{1999}), 3rd ed.
\bibitem{lcls-ii-csr}
\bibinfo{author}{G.~Stupakov}, \bibinfo{title}{\emph{CSR Radiation from BC2 in
  the LCLS-II}}, Preprint LCLSII-TN-15-39, SLAC (\bibinfo{date}{2015}).
\bibitem{nslsII07}
\bibinfo{title}{\emph{National Synchrotron Light Source-II. Preliminary Design
  Report}}, National Synchrotron Light Source-II. Preliminary Design Report.
  http://www.bnl.gov/nsls2/project/PDR, BNL (\bibinfo{date}{2007}).
\bibitem{cur_preprint}
\bibinfo{author}{G.~Stupakov} and \bibinfo{author}{D.~Zhou},
  \bibinfo{title}{\emph{Longitudinal impedance due to coherent undulator
  radiation in a rectangular waveguide}}, Preprint SLAC-PUB-14332, SLAC
  (\bibinfo{date}{2010}).
\bibitem{gradshteyn00r}
\bibinfo{author}{I.~Gradshteyn} and \bibinfo{author}{I.~Ryzhik},
  \bibinfo{title}{\emph{Table of integrals, series, and products}}
  (\bibinfo{publisher}{Academic Press}, \bibinfo{year}{2000}), 6th ed.

\end{thebibliography}

\appendix

%
\section{Calculation of infinite sum in Eq.~\eqref{eq:23}}\label{app:1}
%

The infinite sum in Eq.~\eqref{eq:23} can be written as
    \begin{align}\label{eq:59}
    S
    &=
    \sum_{m=-\infty}^\infty
    (-1)^m
    \frac{1}{\sqrt{q^2+m^2a^2}}
    \exp\left[
    ik
    \sqrt{
    m^2a^2
    +
    q^2
    }
    \right]
    ,
    \end{align}
where $q=c\tau(s,s')=|\vec r_0(s)-\vec r_0(s')|$. We first introduce an infinite sum of the delta functions and add an integration over a continuous variable $t$,
    \begin{align}\label{eq:60}
    S
    &=
    \int_{-\infty}^\infty
    dt
    \sum_{m=-\infty}^\infty
    \delta(t-m)
    \frac{\cos(\pi t)}{\sqrt{q^2+t^2a^2}}
    \exp\left[
    ik
    \sqrt{
    t^2a^2
    +
    q^2
    }
    \right]
    .
    \end{align}
We then use the identity
    \begin{align}\label{eq:61}
    \sum_{m=-\infty}^\infty
    \delta(t-m)
    =
    \sum_{p=-\infty}^\infty
    e^{2\pi ipt}
    \end{align}
to obtain
    \begin{align}\label{eq:62}
    S
    &=
    \int_{-\infty}^\infty
    dt
    \sum_{p=-\infty}^\infty
    e^{2\pi ipt}
    \frac{\cos(\pi t)}{\sqrt{q^2+t^2a^2}}
    \exp\left[
    ik
    \sqrt{
    t^2a^2
    +
    q^2
    }
    \right]
    \nonumber\\
    &=
    \frac{1}{a}
    \sum_{p=-\infty}^\infty
    \int_{-\infty}^\infty
    \frac{dt}{\sqrt{q^2/a^2+t^2}}
    e^{(2p+1)i\pi t}
    \exp\left[
    ika
    \sqrt{
    t^2
    +
    q^2/a^2
    }
    \right]
    .
    \end{align}
This can also be written as
    \begin{align}\label{eq:63}
    S
    &=
    \frac{4}{a}
    \sum_{p=0}^\infty
    \int_{0}^\infty
    dt
    \frac{\cos[(2p+1)\pi t]}{\sqrt{q^2/a^2+t^2}}
    \exp\left[
    ika
    \sqrt{
    t^2
    +
    q^2/a^2
    }
    \right]
    .
    \end{align}
Using the integrals 3876.1 and 3876.2 from Ref.~\cite{gradshteyn00r} we find that
    \begin{align}\label{eq:64}
    \int_0^\infty
    \frac{\cos(px)}{\sqrt{x^2+q^2}}
    \exp(i k\sqrt{x^2+q^2})
    dx
    =
    i
    \frac{\pi}{2}
    H_0^{(1)}(q\sqrt{k^2-p^2})
    ,
    \end{align}
where $H_0^{(1)}$ is the Hankel function of the first kind. Returning now to Eq.~\eqref{eq:63} we finally obtain
    \begin{align}\label{eq:65}
    S
    &=
    \frac{2\pi}{a}
    i
    \sum_{p=0}^\infty
    H_0^{(1)}
    \left(q
    \sqrt{k^2-(2p+1)^2\pi^2/a^2}
    \right)
    .
    \end{align}
Substituting this expression for $S$ (see Eq.~\eqref{eq:59}) into~\eqref{eq:23} gives~\eqref{eq:24}.

%
\section{Bend of finite length with shielding}\label{app:B}
%

\subsection{Contribution $Z_1$ from region $0<z'<z<L$}

A formula for $Z_1$ can be easily obtained from~\eqref{eq:40} if we restore integration over $z$ (that was omitted in~\eqref{eq:40}) and change the integration interval for $z'$ from $(-\infty,z)$ to $(0,z)$,
    \begin{align}\label{eq:66}
    Z_1(k)
    &=
    (i-1)
    \frac{\sqrt{\pi k}}{ac\rho^2}
    \sum_{p=0}^\infty
    \int_{0}^{L}
    dz
    \int_{0}^{z}
    dz'
    \left(
    z
    -
    z'
    \right)^{3/2}
    \exp
    \left(
    -i(z-z')
    \frac{(2p+1)^2\pi^2}{2ka^2}
    \right)
    \nonumber\\
    &\times
    \exp\left(
    -ik
    \frac{1}{24\rho^2}(z-z)'^3
    \right)
    .
    \end{align}
Replacing the integration variables $z'$ and $z$ by $\xi = (z-z')k^{1/3}/24^{1/3}\rho^{2/3}$ and $\tau=z k^{1/3}/24^{1/3}\rho^{2/3}$, respectively, we obtain
    \begin{align}\label{eq:67}
    Z_1
    &=
    (i-1)
    \frac{2^{7/2}3^{7/6}\sqrt{\pi}\rho^{1/3}}{ac k^{2/3}}
    {\sum_{p=0}^\infty}
    \int_{0}^l
    d\tau
    \int_{0}^{\tau}
    d\xi\,
    \xi^{3/2}
    \exp\left[
    -
    i\xi^3
    -
    i\xi Q(2p+1)^2
    \right]
    ,
    \end{align}
where $Q=3^{1/3}\pi^2\rho^{2/3}/a^2 k^{4/3}$ and $l=Lk^{1/3}/24^{1/3}\rho^{2/3}$.  Finally, changing the order of integration allows one to take the integral over $\tau$ giving
    \begin{align}\label{eq:68}
    Z_1
    &=
    (i-1)
    \frac{2^{7/2}3\sqrt{Q}}{c\sqrt{\pi}}
    {\sum_{p=0}^\infty}
    \int_{0}^l
    d\xi\,
    (l-\xi)\xi^{3/2}
    \exp\left[
    -
    i\xi^3
    -
    i\xi Q(2p+1)^2
    \right]
    .
    \end{align}

The parameter $Q$ is the shielding parameter $ka^{3/2}/\rho^{1/2}$ to the power $-4/3$. The parameter $l$ is equal to the ratio of $L$ to the formation length of the radiation with wavenumber $k$.

In the limit $l\gg 1$ the factor $l-\xi$ in~\eqref{eq:68} is replaced by $l$, and the impedance becomes proportional to $L$. In this limit, the impedance per unit length $Z_1/L$ reduces to an expression that is equal to~\eqref{eq:41}. 

\subsection{Contribution $Z_2$ from region $z'<0$ and $L<z$}

A simple geometrical analysis gives the following expressions for $\tau$, $s$ and $s'$ as functions of $z$ and $z'$, 
    \begin{align}\label{eq:69}
    c\tau
    &\approx
    \zeta
    +
    \frac{1}{2\zeta}
    \theta_0^2
    \left(z- \frac{1}{2}L\right)^2
    ,\qquad
    s
    \approx
    (z-L)
    \left(
    1
    +
    \frac{1}{2}
    \theta_0^2
    \right)
    +
    L
    \left(
    1
    +
    \frac{1}{6}
    \theta_0^2
    \right)
    ,
    \qquad
    s'
    =
    z'
    ,
    \end{align}
where $\zeta = z-z'$. We also have $1-\vec \beta(s)\cdot\vec\beta(s') = \frac{1}{2}\theta_0^2$. In these expressions we used the smallness of $\theta_0$ and neglected terms of order higher than $\theta_0^2$.

With account of these relations Eq.~\eqref{eq:29} becomes
    \begin{align}\label{eq:70}
    Z_2
    &=
    (i-1)
    \theta_0^2
    \frac{\sqrt{\pi k}}{ac}
    \sum_{p=0}^\infty
    \int_{L}^{\infty}
    dz
    \int_{-\infty}^{0}
    \frac{dz'}{\sqrt{\zeta}}
    \exp
    \left(
    -i\zeta
    \frac{(2p+1)^2\pi^2}{2ka^2}
    \right)
    \nonumber\\
    &\times
    \exp\left[
    ik
    \left(
    \frac{1}{2\zeta}
    \theta_0^2
    \left(z-L+ \frac{1}{2}L\right)^2
    -
    \frac{1}{2}
    \theta_0^2
    (z-L)
    -
    \frac{1}{6}
    \theta_0^2
    L
    \right)
    \right]
    .
    \end{align}
We now change the integration variables from $z'$ and $z$ to $\xi=\frac{1}{2}k\theta_0^2(z-z')$ and $\tau=\frac{1}{2}k\theta_0^2(z-L)$, respectively, to obtain
    \begin{align}\label{eq:71}
    Z_2
    &=
    (i-1)
    \frac{2^{3/2}\sqrt{\pi}}{kac\theta_0}
    e^{
    -iu/3
    }
    \sum_{p=0}^\infty
    \int_{0}^{\infty}
    d\tau
    e^{
    -i\tau
    }
    \int_{\tau+u}^{\infty}
    \frac{d\xi}{\sqrt{\xi}}
    \exp
    \left(
    -i\xi q
    +
    i
    \frac{1}{\xi}
    \left(\tau + \frac{1}{2}u\right)^2
    \right)
    ,
    \end{align}
where $u = \frac{1}{2}k \theta_0^2 L$ and $q={(2p+1)^2\pi^2}/{k^2a^2\theta_0^2}$. The internal integral can be expressed through the error function $\mathrm{erf}(x)$ and $\mathrm{erfc}(x)=1-\mathrm{erf}(x)$ using the following identity 
    \begin{align}\label{eq:72}
    F(a,b,c)
    &=
    \int_c^{\infty}
    \frac{d\xi}{\sqrt{\xi}}
    \exp
    \left(
    i
    \frac{a}{\xi}
    -
    ib\xi
    \right)
    \nonumber\\&
    =
    \frac{\sqrt{\pi }}{2 \sqrt{i b}}
    e^{-2 \sqrt{a b}}
    \left[
    \mathrm{erf}
    \left(
    \frac{\sqrt{-i a}-\sqrt{i b}c}{\sqrt{c}}\right)+e^{4 \sqrt{ a b}}
    \mathrm{erfc}
    \left(\frac{\sqrt{-i a}+\sqrt{i b}c}{\sqrt{c}}\right)+1
    \right]
    ,
    \end{align}
which gives
    \begin{align}\label{eq:73}
    Z_2
    &=
    (i-1)
    \frac{2^{3/2}\sqrt{\pi}}{kac\theta_0}
    e^{
    -iu/3
    }
    \sum_{p=0}^\infty
    \int_{0}^{\infty}
    d\tau
    e^{
    -i\tau
    }
    F\left(\left(\tau + \frac{1}{2}u\right)^2,q,\tau+u\right)
    .
    \end{align}

If the upper limit of integration over $z$ in Eq.~\eqref{eq:70} is not infinity by some finite value $L+Z$, it is easy to check that Eq.~\eqref{eq:73} is replaced by the following one
    \begin{align}\label{eq:74}
    Z_2
    &=
    (i-1)
    \frac{2^{3/2}\sqrt{\pi}}{kac\theta_0}
    e^{
    -iu/3
    }
    \sum_{p=0}^\infty
    \int_{0}^{u_Z}
    d\tau
    e^{
    -i\tau
    }
    F\left(\left(\tau + \frac{1}{2}u\right)^2,q,\tau+u\right)
    ,
    \end{align}
where $u_Z = \frac{1}{2}k Z \theta_0^2$.

\subsection{Contribution $Z_3$ from region $z'<0$ and $0<z<L$}

In this region, the following expressions for $\tau$, $s$, $s'$ and $1-\vec \beta\cdot\vec\beta'$ are valid, 
    \begin{align}\label{eq:75}
    c\tau
    &\approx
    \zeta
    +
    \frac{1}{2\zeta}
    \frac{1}{4L^2}
    \theta_0^2
    z^4
    ,\qquad
    s
    \approx
    z
    +
    \frac{1}{6L^2}
    \theta_0^2z^3
    ,
    \qquad
    s'
    =
    z'
    ,\qquad
    1-\vec \beta\cdot\vec\beta'
    =
    \frac{1}{2L^2} \theta_0^2z^2
    .
    \end{align}
Substituting them into~\eqref{eq:29} gives
    \begin{align}\label{eq:76}
    Z_3(k)
    &=
    (i-1)
    \frac{\sqrt{\pi k}\theta_0^2}{acL^2}
    \sum_{p=0}^\infty
    \int_{0}^{L}
    z^2
    dz
    \int_{-\infty}^{0}
    \frac{dz'}{\sqrt{\zeta}}
    \exp
    \left(
    -i\zeta
    \frac{(2p+1)^2\pi^2}{2ka^2}
    \right)
    \exp\left[
    ik
    \left(
    \frac{\theta_0^2z^4}{8\zeta L^2}
    -
    \frac{\theta_0^2z^3}{6L^2}
    \right)
    \right]
    .
    \end{align}
We now change the integration variables from from $z'$ and $z$ to $\xi=\frac{1}{2}k\theta_0^2(z-z')$ and $\tau=\frac{1}{2}k\theta_0^2z$, respectively, which gives
    \begin{align}\label{eq:77}
    Z_3(k)
    &\approx
    (i-1)
    \frac{2^{3/2}\sqrt{\pi} }{d^2kac\theta_0}
    \sum_{p=0}^\infty
    \int_0^{u}
    \tau^2
    d\tau
    \exp
    \left(
    -
    \frac{i\tau^3}{3d^2}
    \right)
    \int_{\tau}^\infty
    \frac{d\xi}{\sqrt{\xi}}
    \exp
    \left(
    \frac{i\tau^4}{4\xi d^2}
    -
    iq
    \xi
    \right)
    ,
    \end{align}
where $u = \frac{1}{2}k \theta_0^2 L$, $d= \frac{1}{2}k \theta_0^3\rho$ and $q={(2p+1)^2\pi^2}/{k^2a^2\theta_0^2}$. We then use function $F$ defined by~\eqref{eq:72} to obtain
    \begin{align}\label{eq:78}
    Z_3(k)
    &\approx
    (i-1)
    \frac{2^{3/2}\sqrt{\pi} }{d^2kac\theta_0}
    \sum_{p=0}^\infty
    \int_0^{u}
    \tau^2
    d\tau
    \exp
    \left(
    -
    \frac{i\tau^3}{3d^2}
    \right)
    F\left(\frac{\tau^4}{4 d^2},q,\tau\right)
    .
    \end{align}

\subsection{Contribution $Z_4$ from region $0<z'<L$ and $L<z$}

In this region we have, 
    \begin{align}\label{eq:79}
    c\tau
    &\approx
    \zeta
    \left(
    1
    +
    \frac{1}{2\zeta^2}
    \theta_0^2
    \left(z- \frac{1}{2}L- \frac{1}{2L}z'^2\right)^2
    \right)
    ,\qquad
    s
    \approx
    (z-L)
    \left(
    1
    +
    \frac{1}{2}
    \theta_0^2
    \right)
    +
    L
    \left(
    1
    +
    \frac{1}{6}
    \theta_0^2
    \right)
    ,
    \nonumber\\
    s'
    &=
    z'
    +
    \frac{1}{6}
    \varkappa^2z'^3
    ,\qquad
    1-\vec \beta\cdot\vec\beta' 
    \approx 
    \frac{1}{2}\varkappa^2(L-z')^2
    .
    \end{align}
Using Eq.~\eqref{eq:29} we obtain    
    \begin{align}\label{eq:80}
    Z_4(k)
    &\approx
    (i-1)
    \frac{\sqrt{\pi k} }{ac\rho^2}
    \exp
    \left(
    \frac{1}{3}
    ik\theta_0^2 L
    \right)
    \sum_{p=0}^\infty
    \int_{0}^{L}
    (L-z')^2
    dz'
    \nonumber\\
    &\times
    \int_{L}^\infty
    \frac{dz}{\sqrt{\zeta}}
    \exp
    \left(
    -
    \frac{1}{2}
    ik
    \theta_0^2
    z
    +
    i
    \frac{kz'^3}{6\rho^2}
    +
    i
    \frac{k\theta_0^2}{2\zeta}
    \left(z- \frac{1}{2}L- \frac{z'^2}{2L}\right)^2
    -
    i
    \zeta
    \frac{(2p+1)^2\pi^2}{2ka^2}
    \right)
    .
    \end{align}
Changing  the integration variables from $z$ and $z'$ to $\tau=\frac{1}{2}k\theta_0^2(L-z')$ and $\xi=\frac{1}{2}k\theta_0^2(z-z')$ after simple transformations we obtain
    \begin{align}\label{eq:81}
    Z_4(k)
    &\approx
    (i-1)
    \frac{2^{3/2}\sqrt{\pi} }{d^2kac\theta_0}
    e^{-iu/3}
    \sum_{p=0}^\infty
    \int_{0}^{u}
    \tau^2
    d\tau
    \exp
    \left(
    i\tau
    +
    i
    \frac{(u-\tau)^3}{3d^2}
    -
    i
    \frac{\tau^2}{d}
    \right)
    \int_{\tau}^\infty
    \frac{d\xi}{\sqrt{\xi}}
    \exp
    \left(
    \frac{i\tau^4}{4\xi d^2}
    -
    iq
    \xi
    \right)
    ,
    \end{align}
where  $u = \frac{1}{2}k \theta_0^2 L$, $d= \frac{1}{2}k\theta_0^3\rho$ and $q={(2p+1)^2\pi^2}/{k^2a^2\theta_0^2}$. Again, using~\eqref{eq:72} we reduce $Z_4$ to a one-dimensional integral
    \begin{align}\label{eq:82}
    Z_4(k)
    &\approx
    (i-1)
    \frac{2^{3/2}\sqrt{\pi} }{d^2kac\theta_0}
    e^{-iu/3}
    \sum_{p=0}^\infty
    \int_{0}^{u}
    \tau^2
    d\tau
    \exp
    \left(
    i\tau
    +
    i
    \frac{(u-\tau)^3}{3d^2}
    -
    i
    \frac{\tau^2}{d}
    \right)
    F
    \left(
    \frac{\tau^2}{4d^2}
    ,q,\tau
    \right)
    .
    \end{align}

If the upper limit of integration over $z$ in Eq.~\eqref{eq:80} is not infinity but some finite value $L+Z$, it is easy to check that Eq.~\eqref{eq:82} is replaced by the following one
    \begin{align}\label{eq:83}
    Z_4(k)
    &\approx
    (i-1)
    \frac{2^{3/2}\sqrt{\pi} }{d^2kac\theta_0}
    e^{-iu/3}
    \sum_{p=0}^\infty
    \int_{0}^{u}
    \tau^2
    d\tau
    \exp
    \left(
    i\tau
    +
    i
    \frac{(u-\tau)^3}{3d^2}
    -
    i
    \frac{\tau^2}{d}
    \right)
    \nonumber\\
    &\times
    \left[
    F
    \left(
    \frac{\tau^2}{4d^2}
    ,q,\tau
    \right)
    -
    F
    \left(
    \frac{\tau^2}{4d^2}
    ,q,\tau+u_Z
    \right)
    \right]
    .
    \end{align}
where $u_Z = \frac{1}{2}k \theta_0^2 Z$.

Using Eqs.~\eqref{eq:79} we can now show that the contribution from this region does not converge if one uses Eq.~\eqref{eq:16} (that is the impedance for free space) instead of~\eqref{eq:29}. Indeed, comparing these two expressions we see that, apart from a factor, ~\eqref{eq:16} can be obtained from ~\eqref{eq:29} by omitting the term $    -i\zeta{(2p+1)^2\pi^2}/{2ka^2}$ in the exponential function, replacing $\sqrt{\zeta}\to\zeta$ in the denominator, and dropping the summation over $p$,
    \begin{align}\label{eq:84}
    Z_4^\mathrm{(no\,\,shield)}(k)
    &\propto
    \int_{0}^{L}
    (L-z')^2
    dz'
    \int_{L}^\infty
    \frac{dz}{z-z'}
    \exp
    \left(
    -
    \frac{1}{2}
    ik
    \theta_0^2
    z
    +
    i
    \frac{kz'^3}{6\rho^2}
    +
    i
    \frac{k\theta_0^2}{2(z-z')}
    \left(z- \frac{1}{2}L- \frac{z'^2}{2L}\right)^2
    \right)
    .
    \end{align}
In the limit $z\to\infty$ the exponential function in the integrand tends to a constant value and the integral over $z$ diverges logarithmically at the upper limit.

%
\section{Wiggler of finite length}\label{app:C}
%

%
\subsection{Contribution $Z_1$ from region $0<z'<z<L_w$}
%

In this region Eqs.~\eqref{eq:45} are valid. Substituting them into~\eqref{eq:29} we obtain
    \begin{align}\label{eq:85}
    Z_1(k)
    &=
    (i-1)
    \theta_0^2
    \frac{\sqrt{\pi k}}{ac}
    \sum_{p=0}^\infty
    \int_{0}^{L_w}
    dz
    \int_{0}^{z}
    \frac{dz'}{\sqrt{\zeta}}
    \left(
    \sin (k_wz)
    -
    \sin (k_wz')
    \right)^2
    \exp
    \left(
    -i\zeta
    \frac{(2p+1)^2\pi^2}{2ka^2}
    \right)
    \nonumber\\
    &\times
    \exp\left[
    ik\left(
    \frac{\theta_0^2}{2k_w^2\zeta}
    (\cos (k_wz)-\cos (k_wz'))^2
    -
    \frac{1}{4}\theta_0^2
    \zeta
    +
    \frac{\theta_0^2}{8k_w}
    (
    \sin 2k_wz
    -
    \sin 2k_wz'
    )
    \right)
    \right]
    .
    \end{align}
Using dimensionless variables, $\xi=k_w z$, $\nu = k_w \zeta$ and replacing the integration over $z'$ by integration over $\zeta$ we rewrite~\eqref{eq:85} as
    \begin{align}\label{eq:86}
    Z_1(k)
    &=
    (i-1)
    \theta_0^2
    \frac{\sqrt{\pi k}}{k_w^{3/2}ac}
    \sum_{p=0}^\infty
    \int_{0}^{u}
    d\xi
    \int_{0}^{\xi}
    \frac{d\nu}{\sqrt{\nu}}
    \left(
    \sin \xi
    -
    \sin (\xi-\nu)
    \right)^2
    \exp
    \left(
    -i\nu
    \frac{(2p+1)^2\pi^2}{2kk_wa^2}
    \right)
    \nonumber\\
    &\times
    \exp\left[
    i
    \frac{k\theta_0^2}{k_w}
    \left(
    \frac{1}{2\nu}
    (\cos \xi-\cos (\xi-\nu))^2
    -
    \frac{1}{4}
    \nu
    +
    \frac{1}{8}
    (
    \sin 2\xi
    -
    \sin 2(\xi-\nu)
    )
    \right)
    \right]
    ,
    \end{align}
where $u=k_wL_w=2\pi N_w$.

While direct numerical integration in~\eqref{eq:86} is possible, it is slow due to the oscillating nature of the integrand. The following transformation makes is faster.  We first change the order of integration
    $
    \int_{0}^{u}
    d\xi
    \int_{0}^{\xi}
    {d\nu}
    =
    \int_{0}^{u}
    {d\nu}
    \int_{\nu}^{u}
    d\xi
    $.
We then use the expansion
    \begin{align}\label{eq:88}
    e^{ia\cos\phi}
    =
    \sum_{n=-\infty}^\infty
    J_n(a)
    e^{in(\pi/2-\phi)}
    \end{align}
and rewrite~\eqref{eq:86} as follows using the notation $r=k\theta_0^2/k_w$
    \begin{align}\label{eq:89}
    Z_1(k)
    &=
    (i-1)
    \theta_0^2
    \frac{2\sqrt{\pi k}}{k_w^{3/2} ac}
    \sum_{p=0}^\infty
    \sum_{n=-\infty}^\infty
    \int_{0}^{u}
    \frac{d\nu}{\sqrt{\nu}}
    \int_{\nu}^{u}
    d\xi
    \sin^2\frac{\nu}{2}
    \left(
    1
    +
    \cos\left(2\xi-{\nu}\right)
    \right)
    \exp
    \left(
    -i\nu
    \frac{(2p+1)^2\pi^2}{2kk_wa^2}
    \right)
    \nonumber\\
    &\times
    \exp\left[
    ir
    \left(
    \frac{1}{\nu}
    \sin^2\frac{\nu}{2}
    -
    \frac{1}{4}
    \nu
    \right)
    \right]
    J_n\left(
    \frac{1}{4}
    r
    \sin \nu
    -
    \frac{r}{\nu}
    \sin^2\frac{\nu}{2}
    \right)
    e^{in(\pi/2-2\xi+\nu)}
    .
    \end{align}
Integration over $\xi$ can be carried out with the help of the following
    \begin{align}\label{eq:90}
    &G(\nu,n,u)
    =
    \int_\nu^{u}
    d\xi
    \left(
    1
    +
    \cos\left(2\xi-{\nu}\right)
    \right)
    e^{in(\pi/2-2\xi+\nu)}
    \\
    &=
    \frac{1}{2 n
    \left(n^2-1\right)}
    \left[{i^{n+1} e^{-i \nu  n} \left(\left(n^2-1\right)
   \left(e^{2 i \nu  n}-1\right)+n^2 \cos (\nu )
   \left(e^{2 i \nu  n}-1\right)-i n \sin (\nu )
   \left(e^{2 i \nu  n}+1\right)\right)}
    \right]
    \nonumber
    ,
    \end{align}
where $u=2\pi m$ with $m$ integer. For $n=0$ the integral is $G(\nu,0,u)=u-\nu-\sin\nu$ and for $n=1$ it is equal to
    \begin{align}\label{eq:91}
    G(\nu,\pm 1,u)
    =
    \pm
    \frac{1}{8} 
    e^{-2 i \nu } 
    \left(-4 i e^{2 i \nu } (\nu -u)+4 e^{i \nu }-4 e^{3 i \nu }-e^{4 i \nu }+1\right)
    .
    \end{align}
Hence Eq.~\eqref{eq:89} can be replace by
    \begin{align}\label{eq:92}
    Z_1(k)
    &=
    -
    (1-i)
    \theta_0^2
    \frac{2\sqrt{\pi k}}{k_w^{3/2} ac}
    \sum_{p=0}^\infty
    \sum_{n=-\infty}^\infty
    \int_{0}^{u}
    \frac{d\nu}{\sqrt{\nu}}
    G(\nu,n,u)
    \sin^2\frac{\nu}{2}
    \exp
    \left(
    -i\nu
    \frac{(2p+1)^2\pi^2}{2kk_wa^2}
    \right)
    \nonumber\\
    &\times
    \exp\left[
    ir
    \left(
    \frac{1}{\nu}
    \sin^2\frac{\nu}{2}
    -
    \frac{1}{4}
    \nu
    \right)
    \right]
    J_n\left(
    \frac{1}{4}
    r
    \sin \nu
    -
    \frac{r}{\nu}
    \sin^2\frac{\nu}{2}
    \right)
    ,
    \end{align}
where the double integral is replaced by a sum (over $n$) of single integrals.

%
\subsection{Contribution $Z_2$ from region $-\infty<z'<0$ and $0<z<L_w$}\label{sec:z2_und}
%

In this region we have
    \begin{align}\label{eq:93}
    c\tau
    &\approx
    \zeta
    +
    \frac{\theta_0^2}{2k_w^2\zeta}
    (\cos (k_wz)-1)^2
    \nonumber\\
    1-\vec \beta\cdot\vec\beta'
    &=
    1
    -
    \left(
    1
    -
    \frac{1}{2}
    \theta_0^2
    \sin^2 (k_wz)
    \right)
    =
    \frac{1}{2}
    \theta_0^2
    \left(
    \sin (k_wz)
    \right)^2
    ,
    \end{align}
and
    \begin{align}\label{eq:94}
    c\tau-s(z)+s(z')
    =
    \frac{\theta_0^2}{2k_w^2\zeta}
    (\cos (k_wz)-1)^2
    -
    \frac{1}{4}\theta_0^2
    z
    +
    \frac{\theta_0^2}{8k_w}
    \sin 2k_wz
    .
    \end{align}
Substituting these relations into~\eqref{eq:29} we obtain
    \begin{align}\label{eq:95}
    Z_2(k)
    &=
    (i-1)
    \theta_0^2
    \frac{\sqrt{\pi k}}{ac}
    \sum_{p=0}^\infty
    \int_{0}^{L_w}
    dz
    \int_{-\infty}^{0}
    \frac{dz'}{\sqrt{\zeta}}
    \left(
    \sin (k_wz)
    \right)^2
    \exp
    \left(
    -i\zeta
    \frac{(2p+1)^2\pi^2}{2ka^2}
    \right)
    \nonumber\\
    &\times
    \exp\left[
    ik\left(
    \frac{\theta_0^2}{2k_w^2\zeta}
    (\cos (k_wz)-1)^2
    -
    \frac{1}{4}\theta_0^2
    z
    +
    \frac{\theta_0^2}{8k_w}
    \sin 2k_wz
    \right)
    \right]
    .
    \end{align}
Using dimensionless variables $\xi=k_w z$, $\nu = k_w \zeta$ and replacing integration over $z'$ by integration over $\zeta$ we find
    \begin{align}\label{eq:96}
    Z_2(k)
    &=
    (i-1)
    \theta_0^2
    \frac{\sqrt{\pi k}}{k_w^{3/2}ac}
    \sum_{p=0}^\infty
    \int_{0}^{u}
    d\xi
    \exp\left[
    i
    \frac{k\theta_0^2}{k_w}
    \left(
    -
    \frac{1}{4}
    \xi
    +
    \frac{1}{8}
    \sin 2\xi
    \right)
    \right]
    \left(
    \sin \xi
    \right)^2
    \nonumber\\
    &\times
    \int_\xi^{\infty}
    \frac{d\nu}{\sqrt{\nu}}
    \exp
    \left(
    -i\nu
    \frac{(2p+1)^2\pi^2}{2kk_wa^2}
    \right)
    \exp\left[
    i
    \frac{k\theta_0^2}{k_w}
    \frac{1}{2\nu}
    (\cos \xi-1)^2
    \right]
    .
    \end{align}
where $u=k_wL_w=2\pi N_w$.

%
\subsection{Contribution $Z_3$ from region $0<z'<L_w$ and $L_w<z<\infty$}\label{sec:z3_und}
%

In this region we have
    \begin{align}\label{eq:97}
    c\tau
    &\approx
    \zeta
    +
    \frac{\theta_0^2}{2k_w^2\zeta}
    (1-\cos (k_wz'))^2
    \nonumber\\
    1-\vec \beta\cdot\vec\beta'
    &=
    1
    -
    \left(
    1
    -
    \frac{1}{2}
    \theta_0^2
    \sin^2 (k_wz')
    \right)
    =
    \frac{1}{2}
    \theta_0^2
    \left(
    \sin (k_wz')
    \right)^2
    \nonumber]\\
    s
    &=
    z
    +
    \frac{1}{4}\theta_0^2
    L_w
    ,
    \end{align}
which gives
    \begin{align}\label{eq:98}
    c\tau-s(z)+s(z')
    =
    \frac{\theta_0^2}{2k_w^2\zeta}
    (1-\cos (k_wz'))^2
    -
    \frac{1}{4}\theta_0^2
    L_w
    +
    \frac{1}{4}\theta_0^2
    z'
    -
    \frac{\theta_0^2}{8k_w}
    \sin 2k_wz'
    .
    \end{align}
Substituting this into Eq.~\eqref{eq:29} we find
    \begin{align}\label{eq:99}
    Z_3(k)
    &=
    (i-1)
    \theta_0^2
    \frac{\sqrt{\pi k}}{ac}
    \sum_{p=0}^\infty
    \int_0^{L_w}
    dz'
    \exp\left[
    ik\left(
    -
    \frac{1}{4}\theta_0^2
    L_w
    +
    \frac{1}{4}\theta_0^2
    z'
    -
    \frac{\theta_0^2}{8k_w}
    \sin 2k_wz'
    \right)
    \right]
    \left(
    \sin (k_wz')
    \right)^2
    \nonumber\\
    &\times
    \int_{L_w-z'}^{\infty}
    \frac{d\zeta}{\sqrt{\zeta}}
    \exp
    \left(
    -i\zeta
    \frac{(2p+1)^2\pi^2}{2ka^2}
    \right)
    \exp\left[
    ik
    \frac{\theta_0^2}{2k_w^2\zeta}
    (1-\cos (k_wz'))^2
    \right]
    .
    \end{align}
Using dimensionless variables $\xi=k_w z'$, $\nu = k_w \zeta$ and replacing the integration over $z$ by integration over $\zeta$ we obtain
    \begin{align}\label{eq:100}
    Z_3(k)
    &=
    (i-1)
    \theta_0^2
    \frac{\sqrt{\pi k}}{k_w^{3/2}ac}
    \sum_{p=0}^\infty
    \int_{0}^{u}
    d\xi
    \exp\left[
    i
    \frac{k\theta_0^2}{k_w}
    \left(
    \frac{1}{4}
    (\xi-u)
    -
    \frac{1}{8}
    \sin 2\xi
    \right)
    \right]
    \left(
    \sin \xi
    \right)^2
    \nonumber\\
    &\times
    \int_{u-\xi}^{\infty}
    \frac{d\nu}{\sqrt{\nu}}
    \exp
    \left(
    -i\nu
    \frac{(2p+1)^2\pi^2}{2kk_wa^2}
    \right)
    \exp\left[
    i
    \frac{k\theta_0^2}{k_w}
    \frac{1}{2\nu}
    (\cos \xi-1)^2
    \right]
    ,
    \end{align}
where $u=k_wL_w=2\pi N_w$.

%
\section{Infinitely long wiggler}\label{app:D}
%

To calculate the impedance of an infinitely long wiggler we use Eq.~\eqref{eq:86} that corresponds to the integration over the orbit inside  the wiggler. In this equation, we replace integration over the whole undulator by averaging over one period
    \begin{align}\label{eq:101}
    Z(k)
    &=
    (i-1)
    \theta_0^2
    \frac{\sqrt{\pi k}}{k_w^{3/2}\lambda_w ac}
    \sum_{p=0}^\infty
    \int_{0}^{2\pi}
    d\xi
    \int_{0}^\infty
    \frac{d\nu}{\sqrt{\nu}}
    \left(
    \sin \xi
    -
    \sin (\xi-\nu)
    \right)^2
    \exp
    \left(
    -i\nu
    \frac{(2p+1)^2\pi^2}{2kk_wa^2}
    \right)
    \nonumber\\
    &\times
    \exp\left[
    i
    \frac{k\theta_0^2}{k_w}
    \left(
    \frac{1}{2\nu}
    (\cos \xi-\cos (\xi-\nu))^2
    -
    \frac{1}{4}
    \nu
    +
    \frac{1}{8}
    (
    \sin 2\xi
    -
    \sin 2(\xi-\nu)
    )
    \right)
    \right]
    .
    \end{align}
Using standard trigonometric identities we cast~\eqref{eq:101} into the following form
    \begin{align}\label{eq:102}
    Z(k)
    &=
    (i-1)
    \theta_0^2
    \frac{2\sqrt{\pi k}}{k_w^{3/2}\lambda_w ac}
    \sum_{p=0}^\infty
    \int_{0}^{2\pi}
    d\xi
    \int_{0}^{\infty}
    \frac{d\nu}{\sqrt{\nu}}
    \sin^2\frac{\nu}{2}
    \left(
    1
    +
    \cos\left(2\xi-{\nu}\right)
    \right)
    \exp
    \left(
    -i\nu
    \frac{(2p+1)^2\pi^2}{2kk_wa^2}
    \right)
    \nonumber\\
    &\times
    \exp\left[
    i
    \frac{k\theta_0^2}{k_w}
    \left(
    \frac{1}{\nu}
    \sin^2\frac{\nu}{2}
    \left(
    1
    -
    \cos \left(2\xi-{\nu}\right)
    \right)
    -
    \frac{1}{4}
    \nu
    +
    \frac{1}{4}
    \sin \nu
    \cos (2\xi-\nu)
    \right)
    \right]
    .
    \end{align}
We now change the order of integration and expand the integrand into the series of the Bessel functions using~\eqref{eq:88},
    \begin{align}\label{eq:103}
    Z(k)
    &=
    (i-1)
    \theta_0^2
    \frac{2\sqrt{\pi k}}{k_w^{3/2}\lambda_w ac}
    \sum_{p=0}^\infty
    \sum_{n=-\infty}^\infty
    \int_{0}^{2\pi}
    d\xi
    \int_{0}^{\infty}
    \frac{d\nu}{\sqrt{\nu}}
    \sin^2\frac{\nu}{2}
    \left(
    1
    +
    \cos\left(2\xi-{\nu}\right)
    \right)
    \exp
    \left(
    -i\nu
    \frac{(2p+1)^2\pi^2}{2kk_wa^2}
    \right)
    \nonumber\\
    &\times
    \exp\left[
    ir
    \left(
    \frac{1}{\nu}
    \sin^2\frac{\nu}{2}
    -
    \frac{1}{4}
    \nu
    \right)
    \right]
    J_n\left(
    \frac{1}{4}
    r
    \sin \nu
    -
    \frac{r}{\nu}
    \sin^2\frac{\nu}{2}
    \right)
    e^{in(\pi/2-2\xi+\nu)}
    ,
    \end{align}
where $r=k\theta_0^2/k_w$. Integration over $\xi$ selects $J_1$ and $J_{-1}$,
    \begin{align}\label{eq:104}
    Z(k)
    &=
    (i-1)
    \theta_0^2
    \frac{4\pi\sqrt{\pi k}}{k_w^{3/2}\lambda_w ac}
    \sum_{p=0}^\infty
    \int_{0}^{\infty}
    \frac{d\nu}{\sqrt{\nu}}
    \sin^2\frac{\nu}{2}
    \exp
    \left(
    -i\nu
    \frac{(2p+1)^2\pi^2}{2kk_wa^2}
    \right)
    \nonumber\\
    &\times
    \exp\left[
    ir
    \left(
    \frac{1}{\nu}
    \sin^2\frac{\nu}{2}
    -
    \frac{1}{4}
    \nu
    \right)
    \right]
    \left[
    J_0\left(
    \frac{1}{4}
    r
    \sin \nu
    -
    \frac{r}{\nu}
    \sin^2\frac{\nu}{2}
    \right)
    +
    i
    J_1\left(
    \frac{1}{4}
    r
    \sin \nu
    -
    \frac{r}{\nu}
    \sin^2\frac{\nu}{2}
    \right)
    \right]
    .
    \end{align}
This expression can be integrated and summed numerically. 

It is interesting to consider the limit of low frequencies, $r\ll 1$, and large gaps, $a\to\infty$. In this limit Eq.~\eqref{eq:104} should reproduce the low-frequency results of Section~\ref{sec:VB}. Analysis shows that the main contribution to~\eqref{eq:104} in this limit comes from the region $\nu\ll 1$, and we can simplify the integrand replacing the Bessel functions $J_0\to 1$, $J_1\to 0$, and also
    \begin{align}\label{eq:105}
    \exp\left[
    ir
    \left(
    \frac{1}{\nu}
    \sin^2\frac{\nu}{2}
    -
    \frac{1}{4}
    \nu
    \right)
    \right]
    \to
    \exp\left[
    -ir
    \frac{1}{4}
    \nu
    \right]
    .
    \end{align}
At the same time, due to large $a$, we replace the summation over $p$ by integration
    \begin{align}\label{eq:106}
    \sum_{p=0}^\infty
    \to
    \int_0^\infty
    dp
    .
    \end{align}
As a result, we obtain
    \begin{align}\label{eq:107}
    Z(k)
    &=
    (i-1)
    \theta_0^2
    \frac{4\pi\sqrt{\pi k}}{k_w^{3/2}\lambda_w ac}
    \int_{0}^{\infty}
    \frac{d\nu}{\sqrt{\nu}}
    \sin^2\frac{\nu}{2}
    \exp\left(
    -ir
    \frac{1}{4}
    \nu
    \right)
    \int_0^\infty
    dp
    \exp
    \left(
    -i\nu
    \frac{(2p+1)^2\pi^2}{2kk_wa^2}
    \right)
    \nonumber\\
    &\approx
    i
    \theta_0^2
    \frac{k}{  c}
    \int_{0}^{\infty}
    \frac{d\nu}{\nu}
    \sin^2\frac{\nu}{2}
    \exp\left(
    -ir
    \frac{1}{4}
    \nu
    \right)
    ,
    \end{align}
which reproduces Eqs.~\eqref{eq:48} and~\eqref{eq:49}.

\end{document}